\documentclass{aastex62}
\usepackage{CJK}

\graphicspath{{./}{figures/}}

\received{February 14, 2020}
\revised{March 31, 2020}
\accepted{\today}
\submitjournal{The Astrophysical Journal}

\shorttitle{CND and torus in NGC 1052}
\shortauthors{Kameno et al.}

\begin{document}
\begin{CJK*}{UTF8}{gbsn}
\title{A Massive Molecular Torus inside a Gas-Poor Cirnumnuclear Disk in the Radio Galaxy NGC 1052 Discovered with ALMA}

\correspondingauthor{Seiji Kameno}
\email{seiji.kameno@alma.cl}

\author[0000-0002-5158-0063]{Seiji Kameno}
\affil{Joint ALMA Observatory, Alonso de C\'{o}rdova 3107 Vitacura, Santiago 763-0355, Chile}

\author[0000-0001-7719-274X]{Satoko Sawada-Satoh}
\affil{Graduate School of Sciences and Technology for Innovation, Yamaguchi University, 1677-1 Yoshida, Yamaguchi-shi, Yamaguchi 753-8512, Japan}

\author[0000-0003-4561-1713]{C. M. Violette Impellizzeri}
\affil{Joint ALMA Observatory, Alonso de C\'{o}rdova 3107 Vitacura, Santiago 763-0355, Chile}

\author[0000-0002-8726-7685]{Daniel Espada}
\affil{SKA Organization, Lower Withington, Macclesfield, Cheshire SK11 9DL, UK}

\author[0000-0002-5461-6359]{Naomasa Nakai}
\affil{Department of Physics, School of Science and Technology, Kwansei Gakuin University, 2-1 Gakuen, Sanda, Hyogo 669-1337, Japan}

\author{Hajime Sugai}
\affil{Kavli Institute for the Physics and Mathematics of the Universe, Todai Institutes for Advanced Study, The University of Tokyo (Kavli IPMU, WPI), 5-1-5 Kashiwanoha, Kashiwa, Chiba 277-8583, Japan}

\author[0000-0003-1780-5481]{Yuichi Terashima}
\affil{Department of Physics, Ehime University, Matsuyama 790-8577, Japan}

\author[0000-0002-4052-2394]{Kotaro Kohno}
\affil{Institute of Astronomy, Graduate School of Science, The University of Tokyo, 2-21-1 Osawa, Mitaka, Tokyo 181-0015, Japan}
\affil{Research Center for the Early Universe, Graduate School of Science, The University of Tokyo, 7-3-1 Hongo, Bunkyo-ku, Tokyo 113-0033, Japan}

\author[0000-0002-2419-3068]{Minju Lee}
\affil{Max-Planck-Institut f\"{u}r Extraterrestrische Physik (MPE), Giessenbachstr. 1, D-85748 Garching, Germnay}

\author[0000-0001-9281-2919]{Sergio Mart\'{i}n}
\affil{European Southern Observatory, Alonso de C\'{o}rdova 3107 Vitacura, Santiago 763-0355, Chile}
\affil{Joint ALMA Observatory, Alonso de C\'{o}rdova 3107 Vitacura, Santiago 763-0355, Chile}



\begin{abstract}
We report ALMA observations of NGC 1052 to quest mass accretion in a gas-poor active galactic nucleus (AGN).
We detected CO emission representing a rotating ring-like circumnuclear disk (CND) seen edge-on with the gas mass of $5.3 \times 10^{5}$ M$_{\odot}$.
The CND has smaller gas mass than that in typical Seyfert galaxies with circumnuclear star formation and is too gas-poor to drive mass accretion onto the central engine.
The continuum emission casts molecular absorption features of CO, HCN, HCO$^+$, SO, SO$_2$, CS, CN, and H$_2$O, with H$^{13}$CN and HC$^{15}$N and vibrationally-excited (v$_2 = 1$) HCN and HCO$^+$.
Broader absorption line widths than CND emission line widths imply presence of a geometrically thick molecular torus with a radius of $2.4 \pm 1.3$ pc and a thickness ratio of $0.7 \pm 0.3$.
We estimate the H$_2$ column density of $(3.3 \pm 0.7) \times 10^{25}$ cm$^{-2}$ using H$^{12}$CN, H$^{13}$CN, and HCO$^{+}$ absorption features and adopting abundance ratio of $^{12}$C-to-$^{13}$C and a HCO$^{+}$-to-H$_2$, and derived the torus gas mass of $(1.3 \pm 0.3) \times 10^7$ M$_{\odot}$, which is $\sim 9$\% of the central black-hole mass. The molecular gas in the torus is clumpy with the estimated covering factor of $0.17^{+0.06}_{-0.03}$.
The gas density of clumps inside the torus is inferred to be $(6.4 \pm 1.3) \times 10^7$ cm$^{-3}$, which meets the excitation conditions of H$_2$O maser.
Specific angular momentum in the torus exceeds a flat-rotation curve extrapolated from that of the CND, indicating a Keplerian rotation inside a 14.4-pc sphere of influence.

\end{abstract}





\section{Introduction} \label{sec:intro}
Active galactic nuclei (AGNs) are supposed to be powered by accretion matter onto a super-massive black hole (SMBH).
Mass accretion is the key process for AGN activities and growth of SMBHs.
Overall process of mass accretion still remains unclear to date; what is the accretion matter--gas, dust, or stars? Where does the matter come from? How can the matter lose its angular momentum to accrete?
There are working hypotheses for angular-momentum transfer of accretion matter; (A):Turbulence triggered by a circumnuclear starburst \citep{2002ApJ...566L..21W, 2009ApJ...702...63W} or collisions of clumpy molecular clouds \citep{2004A&A...413..949V} can work on interstellar gas, (B):Radiation from the AGN can extract angular momentum of dust particles via Poynting--Robertson effect \citep{1997ApJ...479L..97U, 1998MNRAS.299.1123U, 2002MNRAS.329..572K}, (C):Dynamical friction of dense star clusters \citep{2001ApJ...562L..19E}, (D):Magnetic avalanche : a magnetized sub-pc-scale rotating disk produces spinning bipolar jets with twisted magnetic fields that extract angular momentum from the disk itself to enhance the accretion rate \citep{1996ApJ...461..115M, 2001Sci...291...84M}, (E): Non-circular motions within a circumnuclear disk (CND) produces successive shocked regions where loss of angular momentum occurs efficiently \citep{2017ApJ...843..136E}.

Observationally, the mass accretion in active galaxies hosting gas-rich (CNDs) seems to be triggered by star formation and supernova explosion \citep{2016ApJ...827...81I}.
It remains unclear whether such a process is also working in gas-poor systems such as radio galaxies where circumnuclear star formation is inactive.

NGC 1052 is one of the best AGNs to investigate mass accretion process in non-starburst environment.
Thanks to the proximity with the angular distance of 17.5 Mpc ($1^{\prime \prime}$ corresponds to 85 pc), sub-arcsecond radio interferometry allows us to investigate gas distribution and kinematics in the pc-scale circumnuclear region.
The radio galaxy is classified as a low luminosity AGN (LLAGN) with the low Eddington ratio of $L_{\rm bol} / L_{\rm Edd} \sim 0.004$ \citep{2002ApJ...579..530W} and the low X-ray luminosity of $L_{\rm 2-10 \ keV} = 4.60 \times 10^{41}$ erg s$^{-1}$ \citep{2009ApJ...698..528B}.
Polarized H$\alpha$ emission with FWHM $\sim 5000$ km s$^{-1}$ infers that the broad line component locates inside an obscuring torus and seen in scattered light, as well as type-2 Seyfert galaxies \citep{1999ApJ...515L..61B}.
\citet{2011MNRAS.411L..21F} claimed presence of young (age $< 7$ Myr) stellar clusters, $> 4^{\prime \prime}$ ($ > 340$ pc) outside of the nucleus, formed in a recent star formation event probably related with a merger event about 1 Gyr ago \citep{1986AJ.....91..791V}, and the estimated rate of 0.01 M$_{\odot}$ yr$^{-1}$ assuming clusters are spread with a similar density over the whole galaxy. This also indicates that NGC 1052 is inactive in ongoing star formation.

Double-sided radio jet structure in sub-pc to kpc scale \citep{2016A&A...593A..47B, 2020AJ....159...14N} with a sub-relativistic bulk speed of $0.26c - 0.53c$ \citep{2003A&A...401..113V, 2013AJ....146..120L, 2019A&A...623A..27B} indicates ongoing AGN activity.
Optical counterpart of the double-sided jet is found as a bipolar [O{\sc III}] $\lambda$5007 emission \citep{2005ApJ...629..131S}.
The viewing angle of the jet is estimated in the range of $50^{\circ} - 72^{\circ}$ \citep{2001PASJ...53..169K, 2003A&A...401..113V, 2004A&A...426..481K}, based on the assumption that the double-sided jet is intrinsically symmetric. The assumption is questioned by \cite{2019A&A...623A..27B} that claims intrinsic asymmetry. Nevertheless, it is well agreed that the jet orientation near the plane of the sky and the torus is seen nearly edge-on.

NGC 1052 exhibits condensation of multi-phase matter in the circumnuclear region.
H$_2$O maser spectrum is broad ($V_{\rm LSR} = 1400$ - 1800 km s$^{-1}$), redshifted with respect to the systemic velocity of 1492 km s$^{-1}$, and variable \citep{2003ApJS..146..249B, 2005ApJ...620..145K}, and its distribution is associated with the jets in the sub-pc scale \citep{2008ApJ...680..191S}.
Free--free absorption (FFA) toward the nucleus and receding jets suggests the presence of a torus composed of dense ($n_{\rm e} \sim 10^5$ cm$^{-3}$) thermal ($T_{\rm e} \sim 10^4$ K) plasma \citep{2001PASJ...53..169K, 2003PASA...20..134K}.
Positions of H$_2$O masers and the FFA plasma coincide, indicating a geometrically thick molecular torus with an ionized inner surface \citep{2008ApJ...680..191S}.
VLBA images of OH absorption \citep{2008evn..confE..33I} also indicate similar distribution to that of FFA and similar velocity profile to that of H$_2$O maser emission. This is consistent with the geometrically thick torus model.
OH, H{\sc I}, CO, HCO$^+$, and HCN absorption features \citep{2003A&A...401..113V, 2004A&A...428..445L} show broad and redshifted profiles similar to what is observed in H$_2$O maser emission.
Since the masing and absorbing gas must locate in the foreground of the nuclear continuum source, the redshifted spectrum implies non-circular motion containing a momentum approaching to the core.
Korean VLBI Network (KVN) observations revealed HCN ($J=1-0$) and HCO$^+$ ($J=1-0$) absorption features toward the jet \citep{2016ApJ...830L...3S, 2019ApJ...872L..21S}.
The absorption features consists of two velocity components at $V_{\rm LSR} = 1655$ and 1720 km s$^{-1}$.
Both components are redshifted with respect to the systemic velocity and appeared toward the western (receding) side of the jet.
This result is consistent with the thick torus model.

In this paper, we present molecular gas distribution and kinematics in NGC 1052 observed with the Atacama Large Millimeter/Submillimeter Array (ALMA).
\S \ref{sec:observations} describes observation log and reduction procedure. \S \ref{sec:results} states fact sheet of acquired images and spectra, and identifies molecular clouds and absorption lines. \S \ref{sec:discussion} consists of discussion about continuum emission (\S \ref{subsec:continuumComponent}), CO emission in the CND (\S \ref{subsec:CND}), characterization of identified absorption lines (\S \ref{subsec:lineID}), estimation of column density (\S \ref{subsec:columnDensity}), and the molecular torus (\S \ref{subsec:TORUS}). \S \ref{subsec:angularMomentum} addresses angular momentum of molecular gas in the CND and the torus. Then \S \ref{sec:conclusions} summarizes our conclusions.

We employ the systemic velocity of $V_{\rm LSR, radio} = 1492$ km s$^{-1}$ based on our measurements (see \S \ref{subsec:CND}). That corresponds to the luminosity distance of $D_L = 17.6$ Mpc, the angular distance of $D_A = 17.5$ Mpc, and the scale of 85 pc arcsec$^{-1}$ under $H_0 = 73$ km s$^{-1}$ Mpc$^{-1}$, $\Omega_M = 0.27$,  $\Omega_{\Lambda} = 0.73$, and $V_{\rm CMB} = 371$ km s$^{-1}$ toward $(l, b) = (264^{\circ}.14, 48^{\circ}.26)$.

\section{Observations and data reduction} \label{sec:observations}
We have conducted ALMA observations (project code: 2013.1.01225.S) consisting of five executions at three different frequency setups to quest molecular emission and absorption features in the CND and torus of NGC 1052. We employed a standard interferometric scheduling block (SB) consisting of pointing correction, system noise measurements, bandpass calibration, and target scans sandwiched by phase calibration scans. The control software chose J0241-0815, which is NGC 1052 itself, as the phase calibrator in some scans and we integrated those scans for the target, too. We used 4 basebands (BBs) of 2-GHz bandwidth in dual linear polarization. The spectral setup with the channel spacing of 2 GHz/128 ch $= 15.625$ MHz and the Hanning window corresponds to the velocity resolutions of 40.8 km s$^{-1}$, 27.3 km s$^{-1}$, and 26.6 km s$^{-1}$ for CO $J=2-1$, CO $J=3-2$, and HCN $J=4-3$ transitions, respectively. The list of observations is shown in table \ref{tab:obslog}.

\begin{deluxetable}{llccccll}
\tablecaption{Observation Log}
\tablehead{
\colhead{SB} & \colhead{ExecBlock} &  \colhead{Date}  &  \colhead{$\nu_{\rm rep}$} & \colhead{$N_{\rm ant}$} & \colhead{On-source integ.} & \colhead{BP Cal} & \colhead{Flux Cal}
 }
 \colnumbers
 \startdata
a\_06 & Xa76868/X2423 & 2015-08-05 & 229.407 & 37 & $22.7 + 3.1$ & J0423-0120, J0238+1636 & J0423-0120 \\
a\_07 & Xa7a216/X1c4c & 2015-08-08 & 344.100 & 41 & $33.8 + 4.4$ & J0238+1636 & J0238+1636 \\
a\_07 & Xa7c533/X1dc7 & 2015-08-11 & 344.100 & 42 & $33.8 + 4.7$ & J0339-0146 & J0339-0146 \\
b\_07 & Xa830fc/X16e2 & 2015-08-16 & 352.767 & 36 & $23.7 + 2.5$ & J0244+0659 & J0244+0659 \\
b\_07 & Xa830fc/X1a2f & 2015-08-16 & 352.767 & 36 & $47.4 + 5.6$ & J0423-0120 & J0238+1636
\enddata
\tablecomments{ 
(1) Scheduling Block (the common prefix, NGC\_1052\_, and the postfix, \_TE, are omitted); (2) Execution block unique identifier (the common prefix, uid://A002/, is omitted);  (3) Observation date; 
(4) Representative frequency (GHz); (5) Number of antennas; 
(6) On-source integration time (minutes), with additional integration as a phase calibrator; (7) Bandpass calibrators; (8) Flux calibrator.
}
\end{deluxetable} \label{tab:obslog}

We used CASA 5.6.0 for calibration and imaging, following standard calibration procedures except phase calibration toward the target source of NGC 1052 itself (i.e. self calibration).
Bandpass and flux calibrators are listed in table \ref{tab:obslog}, assuming the flux densities of 1.129 Jy, 3.110 Jy, 2.051 Jy, 0.882 Jy, 3.391 Jy for J0423-0120 at 230 GHz, J0238+1636 at 350 GHz, J0339-0146 at 350 GHz, J0244+0659 at 346 GHz, and J0238+1636 at 346 GHz, respectively. By comparing flux densities of flux calibrators between different observations and spectral windows, we estimated systematic errors in flux scaling by 4.4\%.
For scheduling blocks of NGC\_1052\_a\_07\_TE and NGC\_1052\_b\_07\_TE, two executions were combined after the calibration processes.

We used natural weighting for synthesis imaging to maximize the signal-to-noise ratio (SNR) targeting absorption spectra toward the compact core.
The CO emission-line images were processed after continuum subtraction in the visibility domain.

\section{Results} \label{sec:results}

\begin{deluxetable}{lrrrrrrr}
\tablecaption{Image performance}
\tablehead{
\colhead{SB} &  \colhead{Beam$_{\rm maj}$}  &  \colhead{Beam$_{\rm min}$} & \colhead{PA} & \colhead{MRS} & \colhead{rms (cont.)} & \colhead{rms (spectral)} & \colhead{T$_{\rm b}$ rms (spectral)}
 }
 \colnumbers
 \startdata
a\_06  & $0^{\prime \prime}.298$ & $0^{\prime \prime}.212$ & $55^{\circ}.4$ & $1^{\prime \prime}.8$ & $11.8$ & $273.9$ & 0.101 \\
a\_07  & $0^{\prime \prime}.209$ & $0^{\prime \prime}.145$ & $58^{\circ}.3$ & $1^{\prime \prime}.5$ & $21.3$ & $277.2$ & 0.094 \\
b\_07  & $0^{\prime \prime}.184$ & $0^{\prime \prime}.145$ & $59^{\circ}.8$ & $1^{\prime \prime}.1$ & $16.6$ & $327.7$ & 0.121
\enddata
\tablecomments{ 
(1) Scheduling Block name (prefix and postfix are omitted in the same way with table \ref{tab:obslog});  (2) - (4) FWHM (major and minor axes) and position angle of the synthesized beam with natural weighting; (5) Maximum recoverable scale ;
(6) Image rms ($\mu$Jy beam$^{-1}$) for continuum; (7) Image rms ($\mu$Jy beam$^{-1}$) for each spectral channel that correspond to the velocity resolutions of 40.8 km s$^{-1}$, 27.3 km s$^{-1}$, and 26.6 km s$^{-1}$; (8) Brightness temperature rms (K) for each spectral channel.
}
\end{deluxetable} \label{tab:imagestatistics}

\begin{figure}[ht]
\begin{center}
\includegraphics[scale=0.65,angle=0]{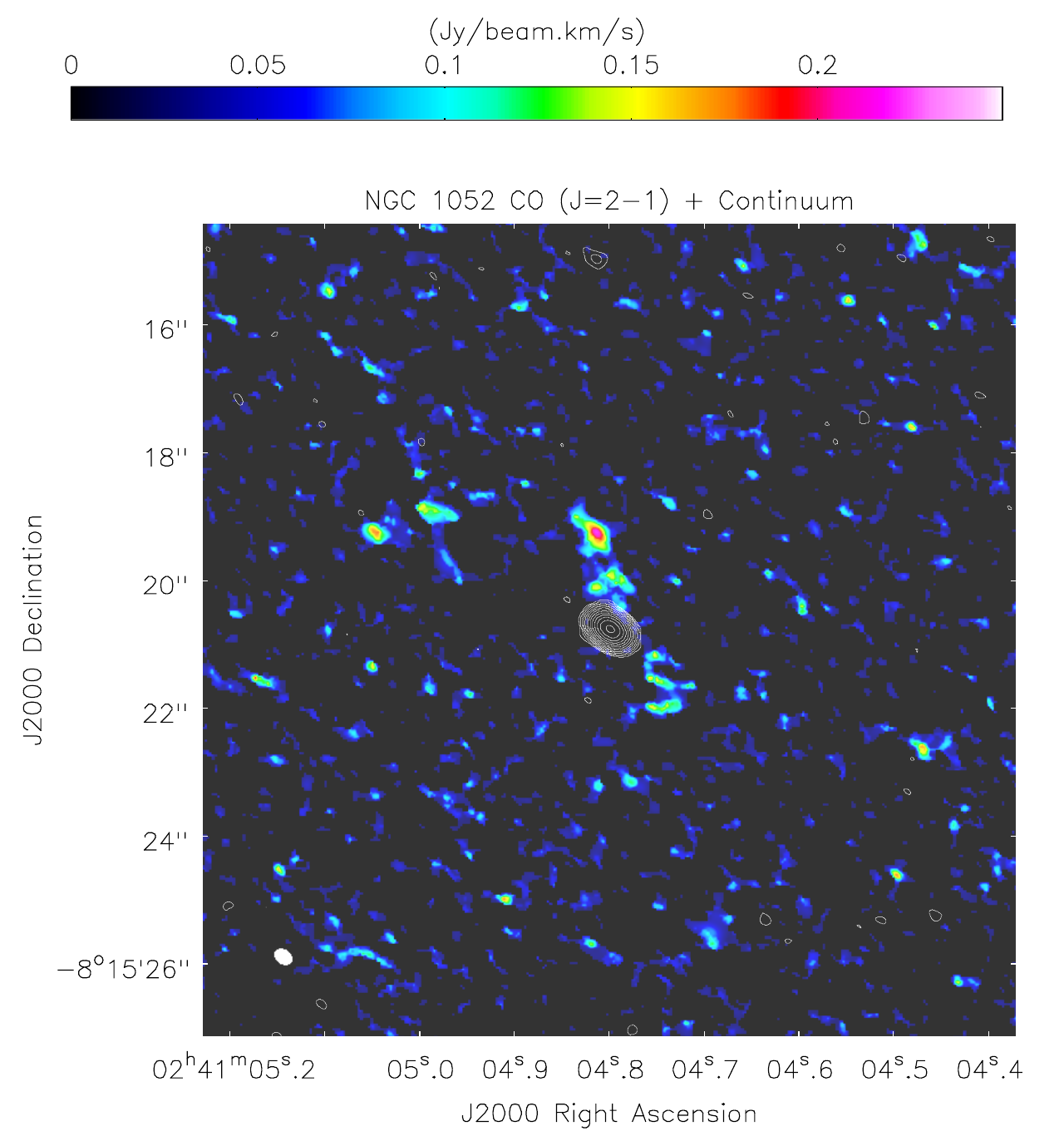}
\includegraphics[scale=0.65,angle=0]{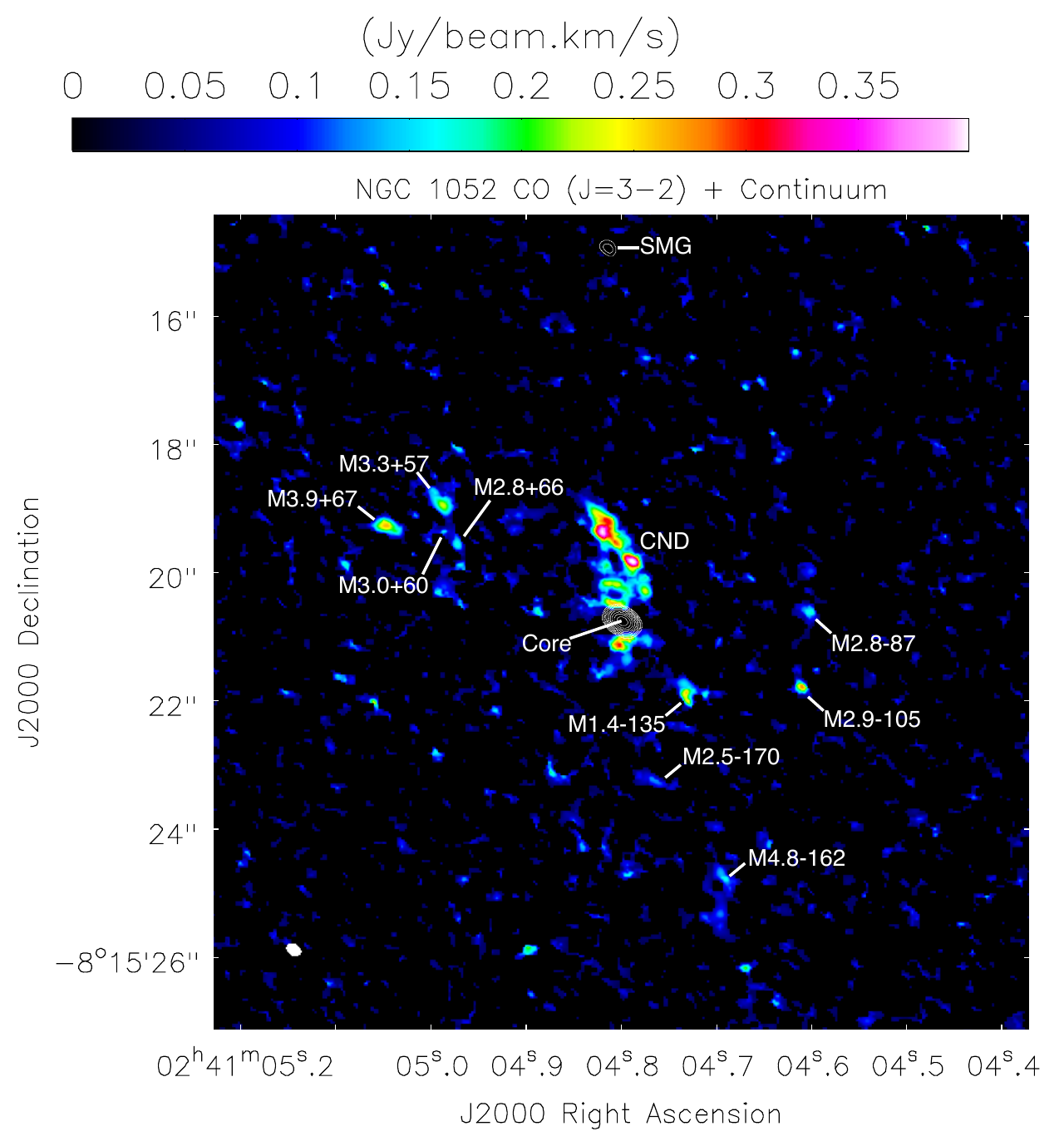}

\caption{CO $J=2-1$ (left) and $J=3-2$ (right) integrated intensity maps of NGC 1052. Continuum level is subtracted in the visibility domain for the CO intensity map. The 222-GHz  and 350-GHz continuum are overlaid on the CO $J=2-1$ and $J=3-2$ images, respectively, by the white contours at increasing powers of 2 times the base of 0.1 mJy beam$^{-1}$. The beam sizes of $0^{\prime \prime}.30 \times 0^{\prime \prime}.21$ in PA$=55^{\circ}$ and $0^{\prime \prime}.21 \times 0^{\prime \prime}.15$ in PA$=58^{\circ}$ are displayed with a white ellipse in the bottom left corner. Labels indicate the continuum core, background submillimeter galaxy (SMG; see \S \ref{subsec:continuum} and \ref{subsec:continuumComponent})., circumnuclear disk (CND), and identified molecular clouds (M) listed in table \ref{tab:MCflux}. \label{fig:Cont+CO}}
\end{center}
\end{figure}

Basic parameters of the resulting images are listed in table \ref{tab:imagestatistics}. Continuum and CO emission-line images are shown in figure \ref{fig:Cont+CO}. Channel maps of CO emission lines with a 45-km s$^{-1}$ bin are presented in figures \ref{fig:CO21ChanMap} and \ref{fig:CO32ChanMap}.

\subsection{Continuum} \label{subsec:continuum}
The continuum emission of NGC 1052 is detected as an unresolved point-like source with the flux densities of $0.487 \pm 0.02$ Jy and $0.442 \pm 0.02$ Jy at 222 GHz and 350 GHz, respectively.
No significant extended feature, such as dust or jet component, has been detected, with the upper limit of 35 $\mu$Jy beam$^{-1}$ and 50 $\mu$Jy beam$^{-1}$ and the maximum recoverable scale of $1^{\prime \prime}.8$ and $1^{\prime \prime}.1$ at 222 GHz and 350 GHz, respectively (see table \ref{tab:imagestatistics}).

We identified an extra unresolved component $\sim 6^{\prime \prime}$ to the north of the core.
The position of the north component is RA=02h41m04s.814 and Dec=$08^{\circ}15^{\prime}14^{\prime \prime}.94$ (J2000 ICRS).
The position uncertainty is $\sim 0^{\prime \prime}.02$, estimated using variance of relative position to the core of NGC 1052 among multiple executions listed in table \ref{tab:obslog}.
The flux densities of the north component is $0.25 \pm 0.01$ mJy and $1.79 \pm 0.07$ mJy at 222 GHz and 350 GHz, respectively.
The north component is almost unresolved with the synthesized beam in table \ref{tab:obslog}, and the difference between the peak flux density of Gaussian fit and the integrated flux density is less than 17\%.

%


\subsection{CO emission} \label{subsec:COemission}

We detected CO $J=2-1$ and $J=3-2$ emission lines in the LSR velocity range of 1160 -- 1800 km s$^{-1}$.
We identified a circumunuclear disk (CND) feature and 9 molecular clouds in the channel maps (see Appendix \ref{appendix:channmap}) with the criteria: (1) peak intensity of CO ($J=3-2$) emission exceeds a threshold of $1.94$ mJy beam$^{-1}$, 7 times as high as the spectral rms to assure $>99$\% confidence for $512 \times 512$ pixels and 128 channels, and (2) the flux density of CO ($J=2-1$) exceeds $0.55 \sqrt{N_{\rm pic}}$ mJy for 95\% confidence, where $N_{\rm pic}$ is the number of connected pixels that meets criterion (1).
These molecular clouds are listed in table \ref{tab:MCflux}.

Figure \ref{fig:Cont+CO} shows the integrated CO intensity maps, together with the continuum component in white contours and labels of identified molecular clouds.
We estimated integrated CO line luminosities, $L^{\prime}_{\rm CO}$, toward each molecular cloud and the CND structure using the conversion formula described by \cite{2005ARA&A..43..677S},
\begin{equation}
L^{\prime}_{\rm CO} = 3.25 \times 10^7 S_{\rm CO} \Delta v \nu^{-2}_{\rm obs} D^2_L (1 + z)^{-3} \  {\rm K \ km \ s}^{-1}.
\end{equation}
These values are listed in table \ref{tab:MCflux}, too.

\begin{figure}[ht]
\begin{center}
\includegraphics[scale=0.65,angle=0]{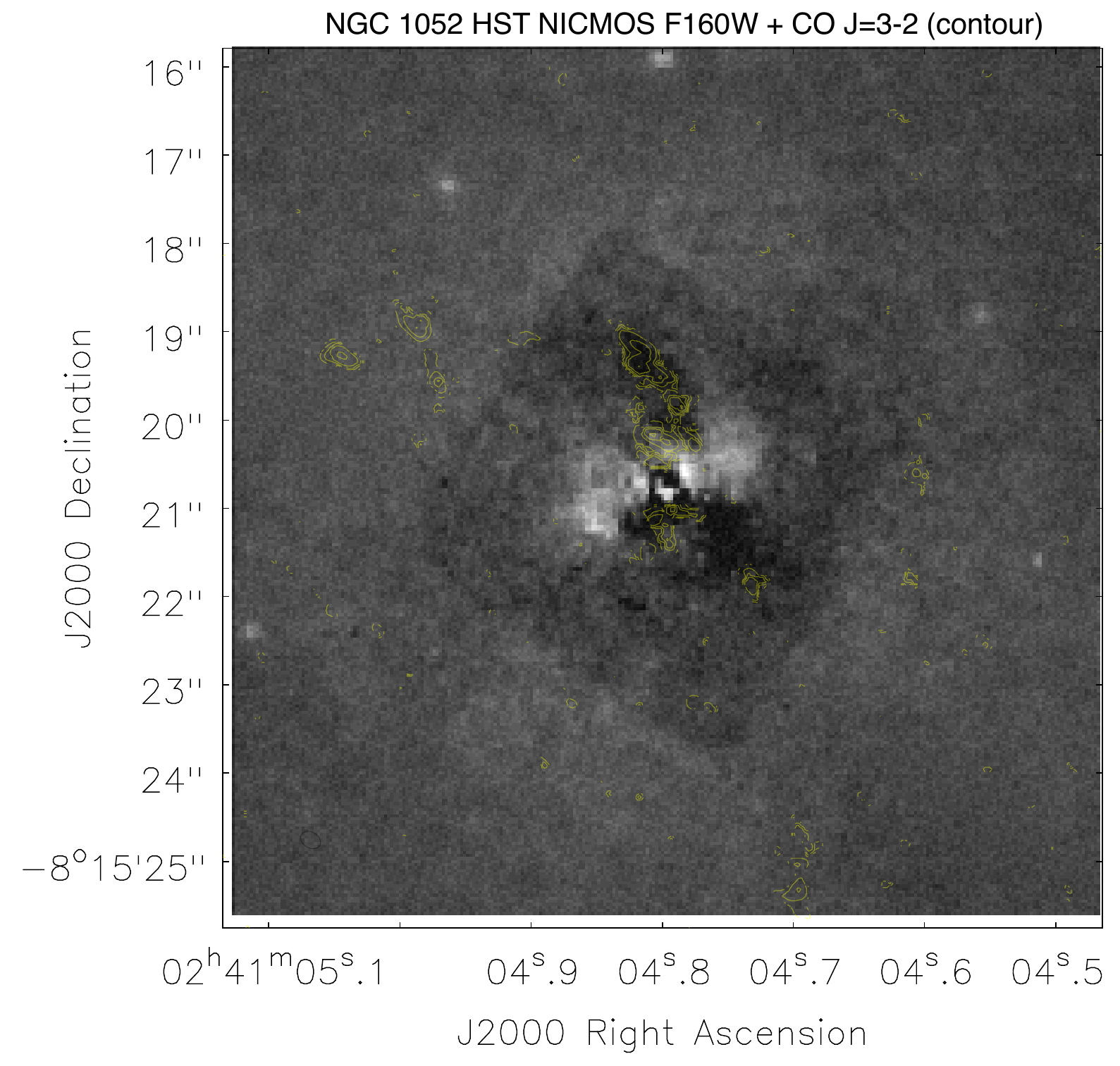}
\caption{CO $J=3-2$ integrated intensity map in yellow contours overlaid on the HST NICMOS F160W image in gray scale \citep{2001AJ....122..653R}. The HST image shows residuals after subtracting spheroid component and the gray scale represents positive values in white, and negative values in dark, which is inverted from the original. The CND with the CO emission coincides the dust absorption.\label{fig:HST+CO}}
\end{center}
\end{figure}

\begin{figure}[ht]
\begin{center}
\includegraphics[scale=0.55,angle=0]{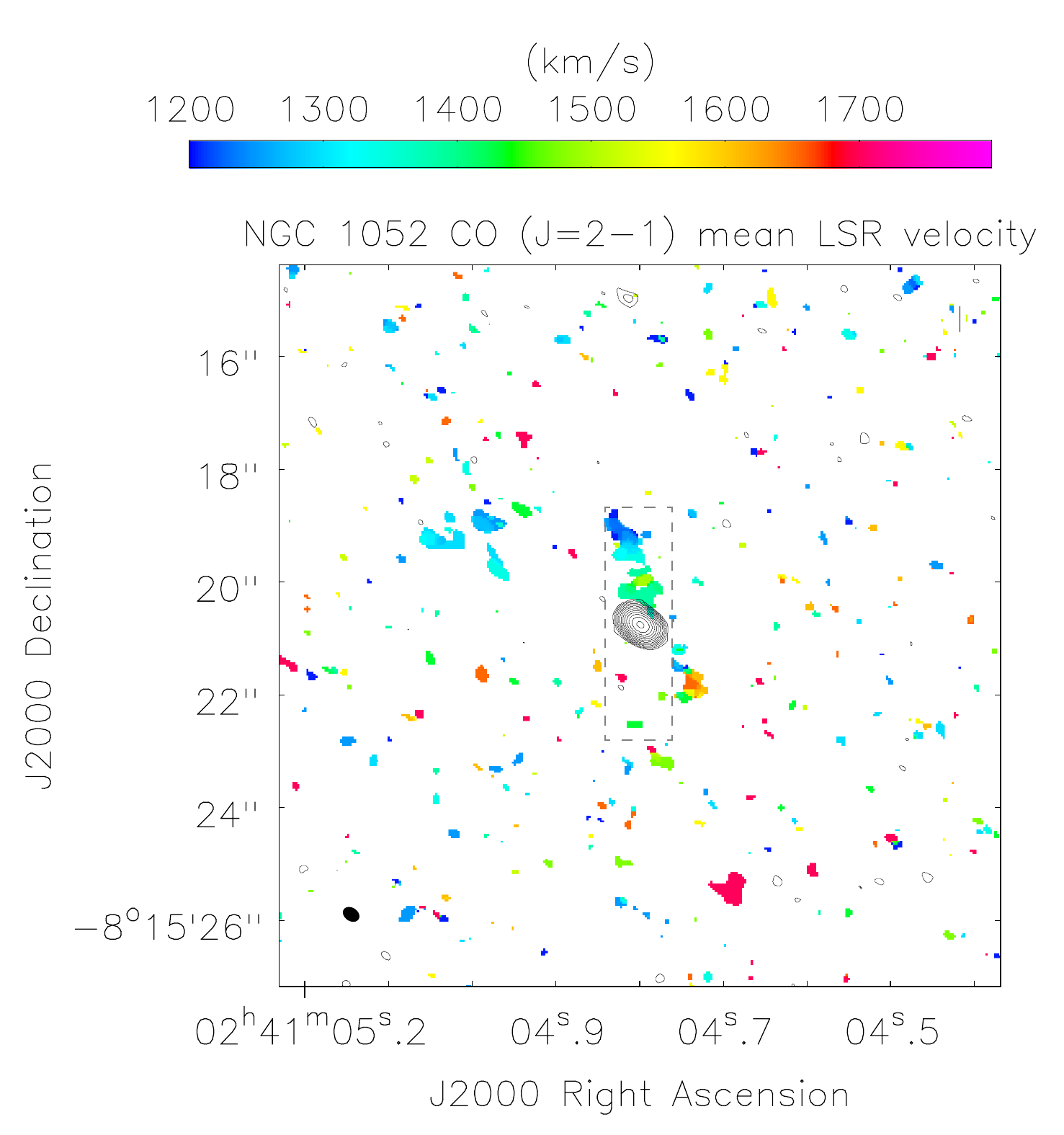}
\includegraphics[scale=0.55,angle=0]{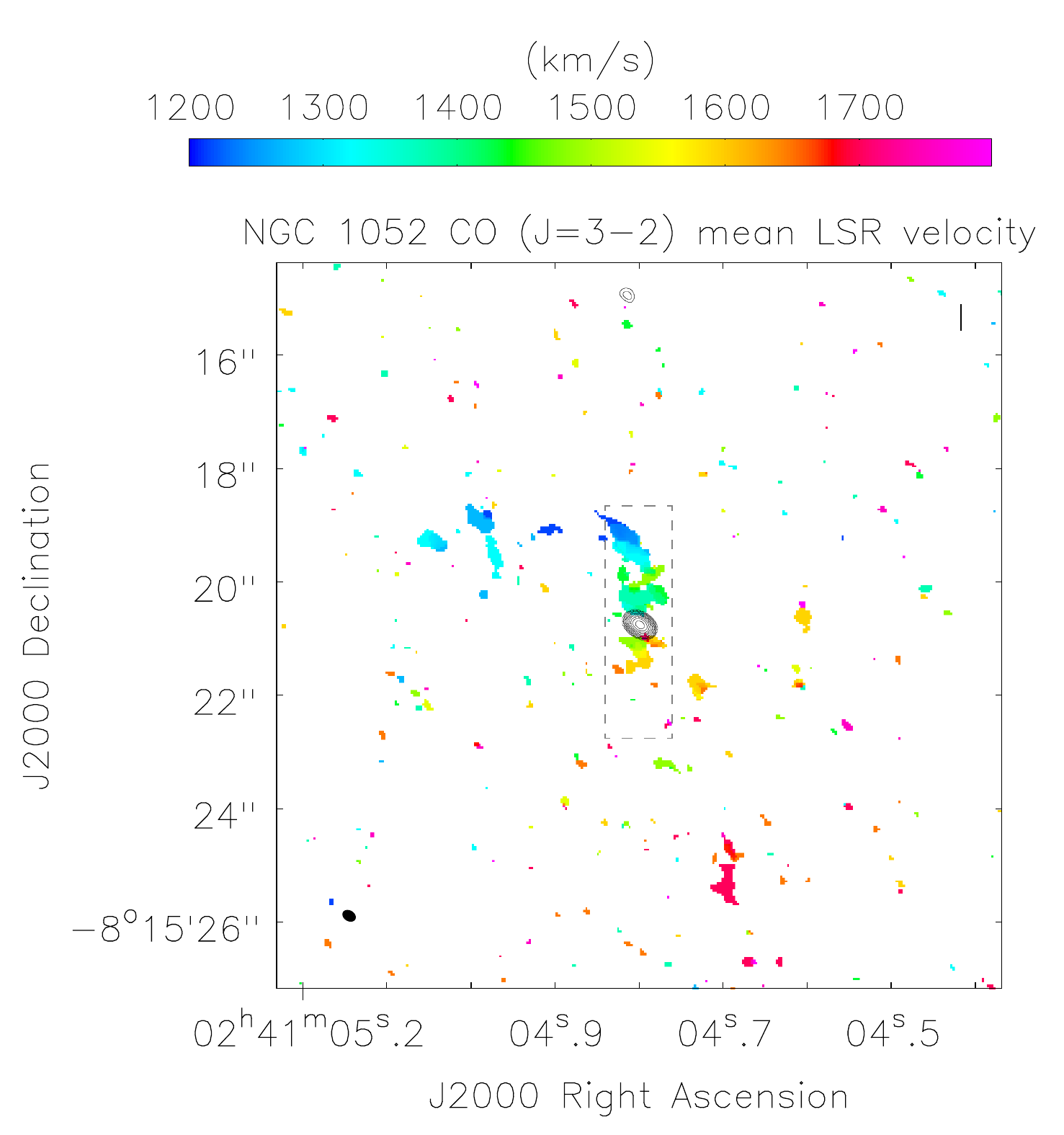}
\caption{Moment-1 (mean LSR velocity) maps of CO $J=2-1$ (left) and $J=3-2$ (right).
We clipped the channel map (figures \ref{fig:CO21ChanMap} and \ref{fig:CO32ChanMap}) by the intensity threshold of 0.52 mJy beam$^{-1}$ and 0.72 mJy beam$^{-1}$, which corresponds to $2\times$ image r.m.s., to generate the moment-1 maps.
The dashed-line rectangles indicate the cut for the position--velocity diagrams shown in figure \ref{fig:COPV}. \label{fig:COvelocity}}
\end{center}
\end{figure}

The CND is identified as a continuous molecular cloud with the largest extent straddling the nucleus. That exhibits velocity gradient along its major axis in the north-south direction.
Figure \ref{fig:COvelocity} shows the moment-1 (mean velocity) maps of CO $J=2-1$ and $J=3-2$ emission lines.
The extent of the CND feature is $1^{\prime \prime}.8$ (153 pc) and $0^{\prime \prime}.9$ (77 pc) to the north and south, respectively, with respect to the nucleus position.
The maximum width along the minor axis is $0^{\prime \prime}.7$ (60 pc) in E-W direction.
The CND structure coincides with the dust absorption feature identified in the Hubble Space Telescope (HST) 1.6 $\mu$m image \citep{2001AJ....122..653R} as shown in figure \ref{fig:HST+CO}.

\begin{deluxetable}{lrrrrrrr}
\tablecaption{Integrated flux densities of molecular clouds identified}
\tablehead{
\colhead{Label} & \colhead{$V_{\rm peak}$} & \colhead{$\Delta \alpha$} & \colhead{$\Delta \delta$} & \colhead{$S_{\rm CO 2-1}$} & \colhead{$S_{\rm CO 3-2}$} & $L^{\prime}_{\rm CO 2-1}$ & $L^{\prime}_{\rm CO 3-2}$\\
\colhead{} & \colhead{(km s$^{-1}$)} & \colhead{($\prime \prime$)} & \colhead{($\prime \prime$)} & \colhead{(Jy km s$^{-1}$)} & \colhead{(Jy km s$^{-1}$)} & \colhead{($10^3$ K km s$^{-1}$ pc$^2$)} & \colhead{($10^3$ K km s$^{-1}$ pc$^2$)}
}
 \colnumbers
 \startdata
M$3.3+57$  & 1275.3 & 2.80 & 1.80 & $0.239 \pm 0.028$ & $0.663 \pm 0.043$ & $44.4 \pm 5.2$ & $54.8 \pm 3.55$ \\ 
M$3.9+67$  & 1302.4 & 3.64 & 1.52 & $0.283 \pm 0.024$ & $0.627 \pm 0.055$ & $52.6 \pm 4.5$ & $51.8 \pm 4.54$ \\ 
M$2.8+66$  & 1315.9 & 2.56 & 1.16 & $0.065 \pm 0.010$ & $0.205 \pm 0.019$ & $12.1 \pm 1.9$ & $16.9 \pm 1.57$ \\ 
M$3.0+60$  & 1315.9 & 2.56 & 1.16 & $0.006 \pm 0.003$ & $0.030 \pm 0.005$ & $ 1.1 \pm 0.56$ & $ 2.5 \pm 0.41$ \\ 
M$2.5-170$ & 1468.3 &-0.44 &-2.44 & $0.073 \pm 0.010$ & $0.169 \pm 0.019$ & $13.6 \pm 1.9$ & $14.0 \pm 1.57$ \\ 
M$2.8-87$  & 1586.8 &-2.76 & 0.12 & $0.009 \pm 0.003$ & $0.054 \pm 0.007$ & $ 1.7 \pm 0.56$ & $ 4.5 \pm 0.58$ \\ 
M$2.9-105$ & 1532.6 &-2.76 &-0.76 & $0.013 \pm 0.005$ & $0.052 \pm 0.009$ & $ 2.4 \pm 0.93$ & $ 4.3 \pm 0.74$ \\ 
M$1.4-135$ & 1613.9 &-1.00 &-1.00 & $0.100 \pm 0.011$ & $0.417 \pm 0.035$ & $18.6 \pm 2.0$ & $34.4 \pm 2.9$ \\ 
M$4.8-162$ & 1712.1 &-1.52 &-4.56 & $0.079 \pm 0.013$ & $0.299 \pm 0.027$ & $14.7 \pm 2.4$ & $24.7 \pm 2.2$ \\ 
CND        & 1356.5 & -    & -    & $1.553 \pm 0.140$ & $7.960 \pm 0.335$ & $288.6\pm 26$& $657.6\pm 28$ \\
\enddata
\tablecomments{ 
(1) Molecular cloud ID consisting of angular distance (arcsec) and position angle (degree) with respect to the nucleus;  (2) LSR velocity at the spectral peak (km s$^{-1}$); (3-4) Position of the peak (except CND) with respect to the nucleus; (5-6) Integrated flux density of CO $J=2-1$ and $J=3-2$ transitions (Jy km s$^{-1}$); (7-8) Luminosity of CO $J=2-1$ and $J=3-2$ transitions (K km s$^{-1}$ pc$^2$).
}
\end{deluxetable} \label{tab:MCflux}

\subsection{Absorption features} \label{subsec:absorption}
\begin{figure}[ht]
\begin{center}
\includegraphics[scale=0.5,angle=0]{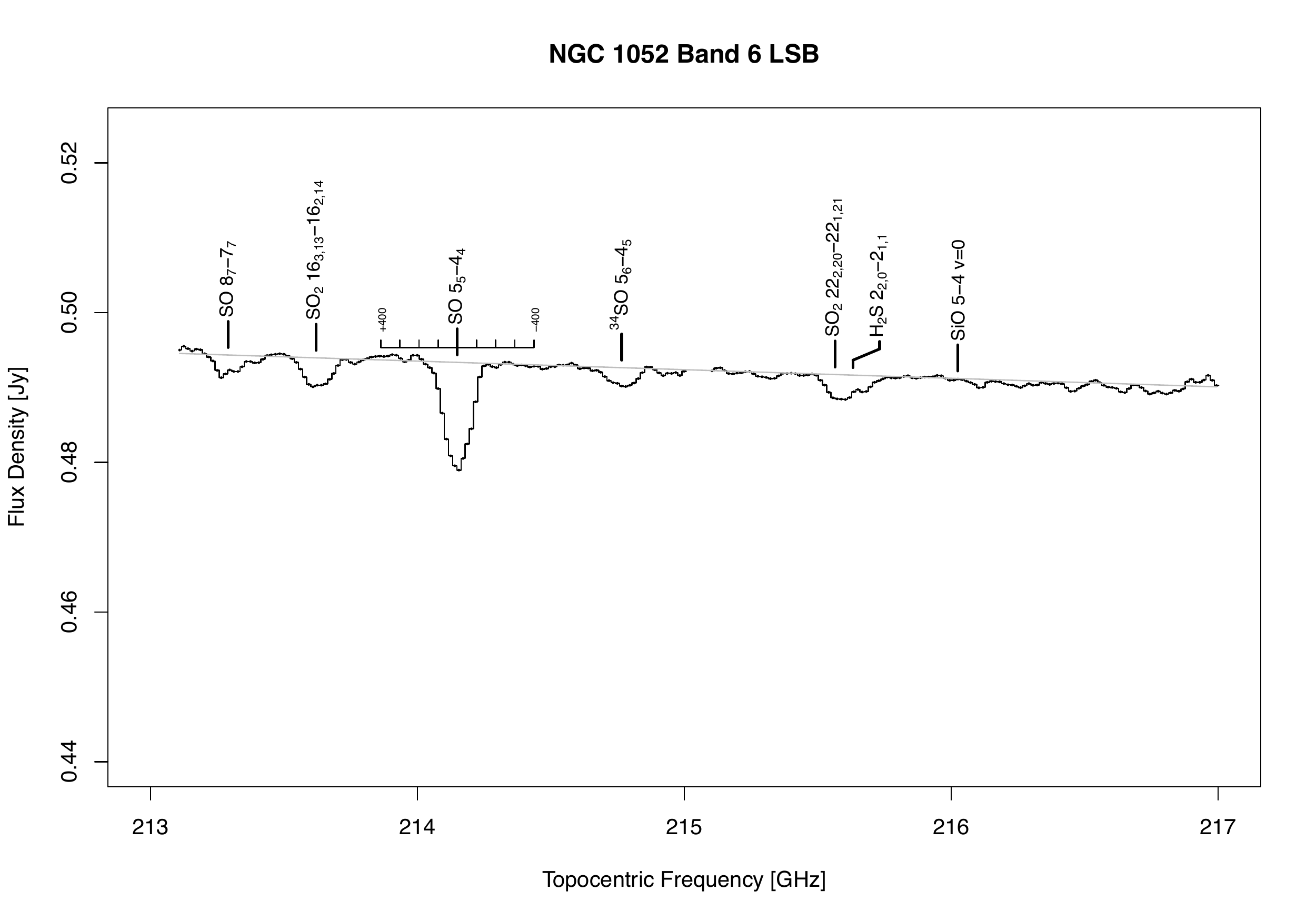}

\includegraphics[scale=0.5,angle=0]{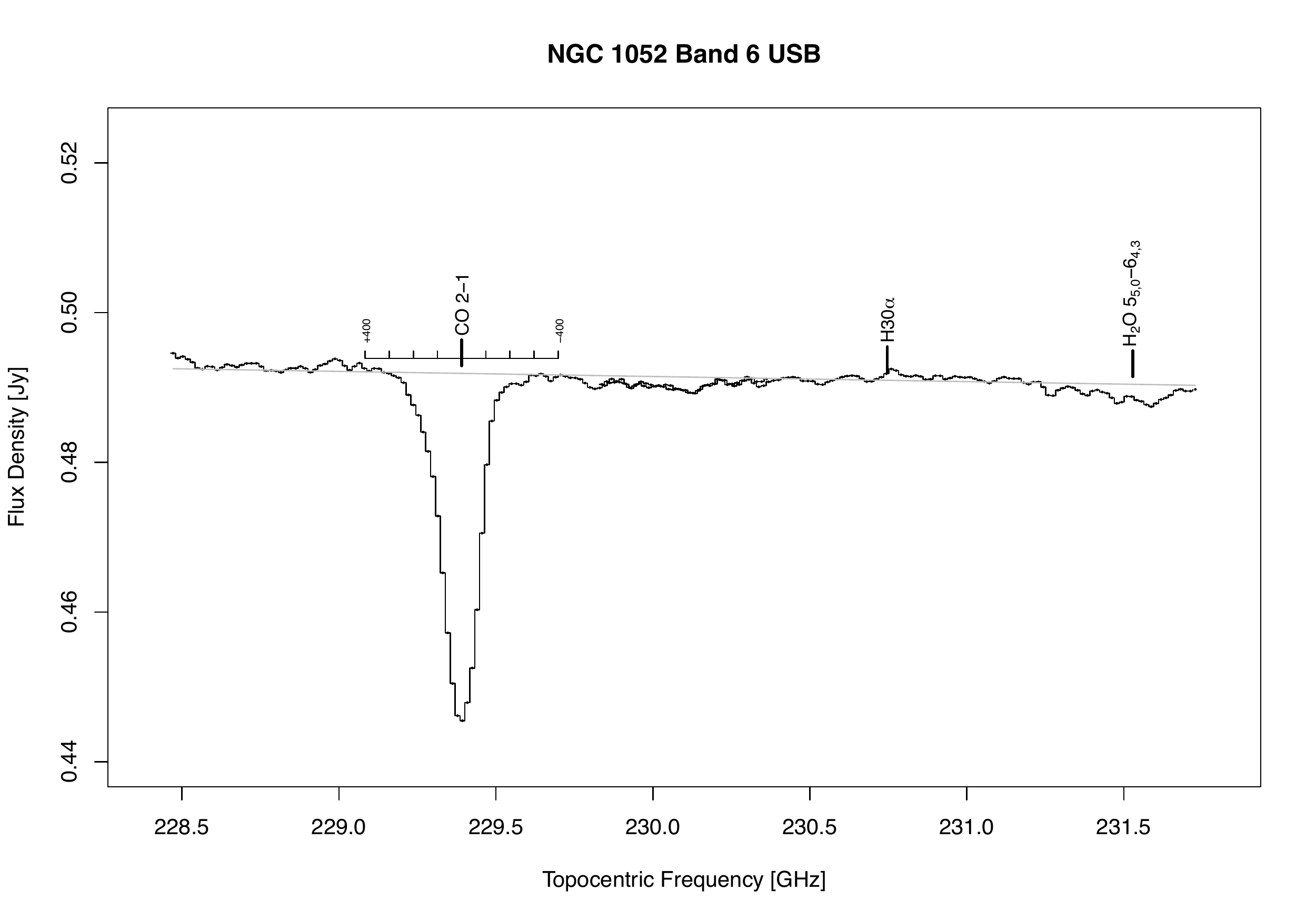}
\caption{Band-6 spectrum toward the NGC 1052 core in topocentric frequency. The gray solid line indicates power-law continuum level inferred from line-free channels. Candidate line species are labelled at the systemic velocity of 1492 km s$^{-1}$. Velocity scales in $\pm 400$ km s$^{-1}$ with respect to the systemic velocity are attached on CO $J=2-1$ and SO $J_N = 5_5 - 4_4$ features. The line identification will be discussed in \S \ref{subsec:lineID}. Frequency coverages of two spectral windows overlap in 229.8 GHz -- 230.4 GHz. There might exist a bandpass residual due to the telluric O$_3$ $J_{Ka, Kc} = 16_{1,15} - 16_{0,16}$ attenuation at 231.28 GHz.\label{fig:B6Spectrum}}
\end{center}
\end{figure}

\begin{figure}[ht]
\begin{center}
\includegraphics[scale=0.7,angle=0]{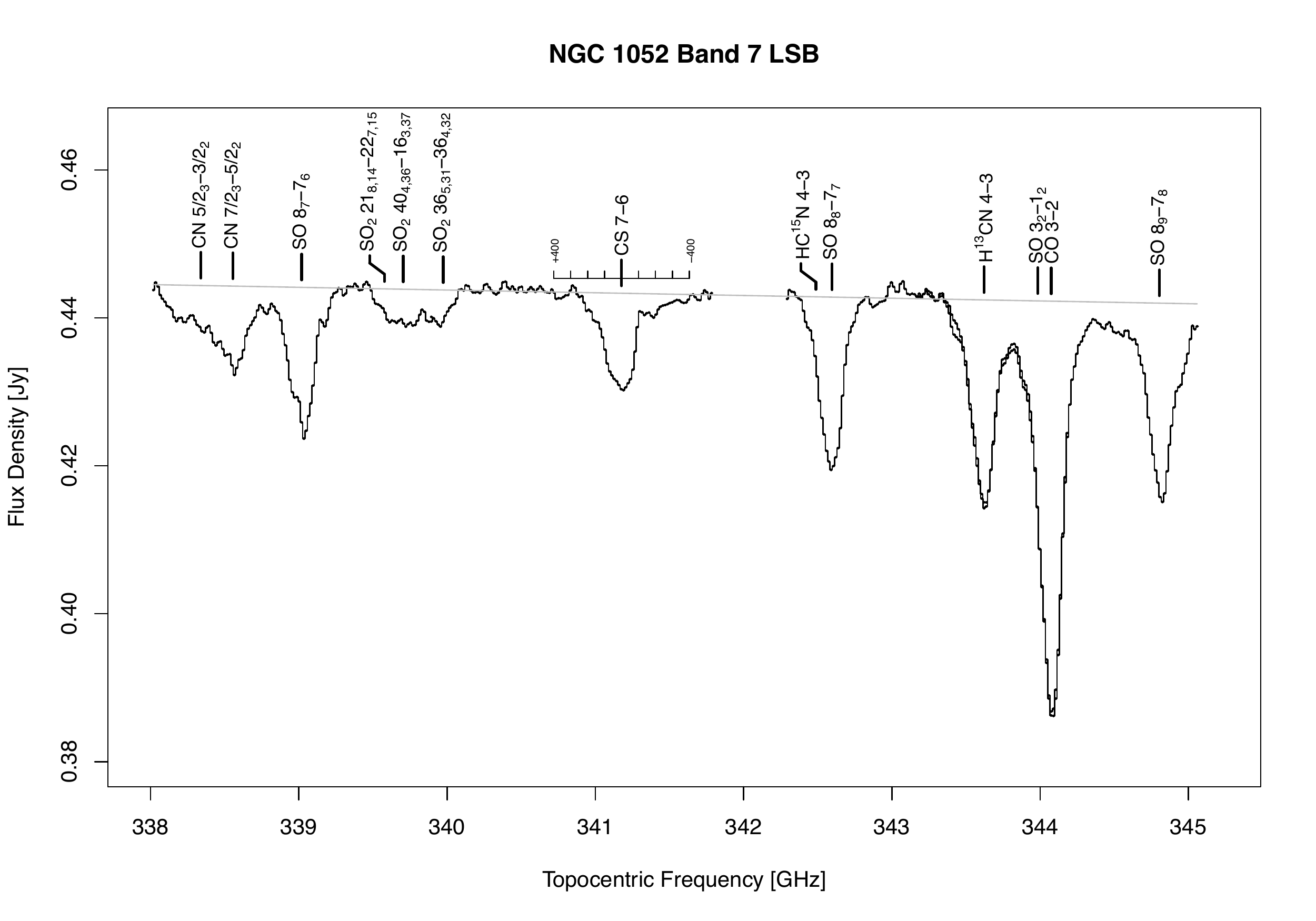}
\caption{Band-7 lower-sideband spectrum toward the NGC 1052 core as well as figure \ref{fig:B6Spectrum}. A velocity scale is attached on CS $J=7-6$ feature. Two spectral windows overlap in the frequency range of 343.2 GHz -- 344.2 GHz.\label{fig:B7LSBSpectrum}}
\end{center}
\end{figure}

\begin{figure}[ht]
\begin{center}
\includegraphics[scale=0.7,angle=0]{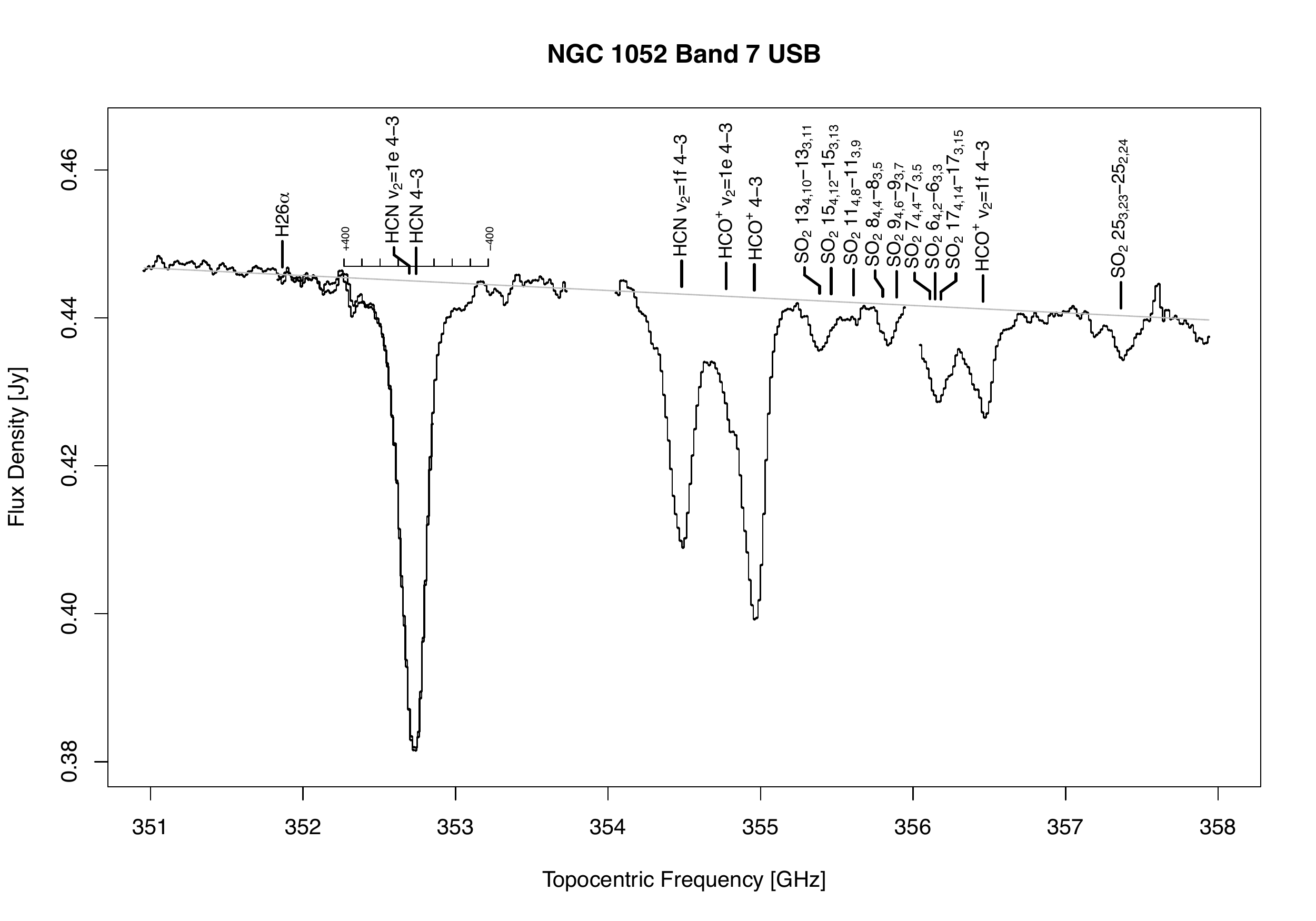}
\caption{Band-7 upper-sideband spectrum toward the NGC 1052 core as well as figure \ref{fig:B7LSBSpectrum}. A velocity scale is attached on HCN $J=4-3$ feature. Two spectral windows overlap in the frequency range of 351.8 GHz -- 352.8 GHz. The feature at 357.63 GHz coincides the telluric O$_3$ $J_{Ka, Kc} = 20_{1,19} - 19_{2,18}$ attenuation.\label{fig:B7USBSpectrum}}
\end{center}
\end{figure}

\begin{deluxetable}{llrrrrr}
\tablecaption{Identified absorption lines}
\tablehead{
\colhead{Species} & \colhead{Transition} & \colhead{$\nu_{\rm rest}$} & \colhead{$\tau_{\rm max}$} & \colhead{LSR velocity} & \colhead{FWHM}          & EW \\
\colhead{}        & \colhead{}           & \colhead{(GHz)}            & \colhead{}                 & \colhead{(km s$^{-1}$)}& \colhead{(km s$^{-1}$)} & \colhead{(km s$^{-1}$)}
}
 \colnumbers
 \startdata
CO        & $J=2-1$           & 230.538 & $0.0993 \pm 0.0017$ & $1496.7 \pm 1.4$ & $166.5 \pm 3.4$ & $17.6 \pm 0.5$ \\
          & $J=3-2$           & 345.796 & $0.1284 \pm 0.0017$ & $1494.7 \pm 1.1$ & $168.8 \pm 2.8$ & $23.1 \pm 0.5$ \\
SO        & $J_N = 8_7 - 7_7$ & 214.357 & $0.0053 \pm 0.0006$ & $1486.4 \pm 8.2$ & $150.8 \pm 19$& $0.9 \pm 0.1$ \\
          & $J_N = 5_5 - 4_4$ & 215.221 & $0.0303 \pm 0.0006$ & $1490.5 \pm 1.4$ & $147.4 \pm 3.3$ & $4.7 \pm 0.1$ \\
          & $J_N = 8_7 - 7_6$ & 340.714 & $0.0417 \pm 0.0017$ & $1489.2 \pm 3.7$ & $182.7 \pm 8.7$ & $8.1 \pm 0.5$ \\
          & $J_N = 8_8 - 7_7$ & 344.311 & $0.0526 \pm 0.0024$ & $1490.5 \pm 4.6$ & $141.1 \pm 5.9$ & $7.9 \pm 0.5$ \\
          & $J_N = 8_9 - 7_8$ & 346.528 & $0.0551 \pm 0.0018$ & $1479.4 \pm 3.2$ & $200.0 \pm 7.6$ & $11.7 \pm 0.6$ \\
$^{34}$SO & $J_N = 5_6 - 4_5$ & 215.839 & $0.0052 \pm 0.0004$ & $1492.0 \pm 5.8$ & $166.8 \pm 14$& $0.9 \pm 0.1$ \\
SO$_2$    & $J_{Ka, Kc} = 16_{3,13} - 16_{2,14}$ & 214.689 & $0.0083 \pm 0.0004$ & $1484.9 \pm 3.6$ & $151.0 \pm 8.5$ & $1.3 \pm 0.1$ \\
          & $J_{Ka, Kc} = 22_{2,20} - 22_{1,21}$ & 216.643 & $0.0067 \pm 0.0003$ & $1438.6 \pm 4.9$ & $211.2 \pm 12$& $1.5 \pm 0.1$ \\
          & $J_{Ka, Kc} = 21_{8,14} - 22_{7,15}$ & 341.276 & $0.0085 \pm 0.0006$ & $1459.4 \pm 5.4$ & $135.3 \pm 5.2$ & $1.2 \pm 0.1$ \\
          & $J_{Ka, Kc} = 40_{4,36} - 40_{3,37}$ & 341.403 & $0.0097 \pm 0.0006$ & $1443.7 \pm 5.6$ & $135.3 \pm 5.2$ & $1.4 \pm 0.1$ \\
          & $J_{Ka, Kc} = 36_{5,31} - 36_{4,32}$ & 341.674 & $0.0105 \pm 0.0004$ & $1517.5 \pm 3.1$ & $135.3 \pm 5.2$ & $1.5 \pm 0.1$ \\
          & $J_{Ka, Kc} = 13_{4,10} - 13_{3,11}$ & 357.165 & $0.0148 \pm 0.0004$ & $1489.7 \pm 2.2$ & $133.1 \pm 4.7$ & $2.1 \pm 0.1$ \\
          & $J_{Ka, Kc} = 15_{4,12} - 15_{3,13}$ & 357.241 & $\updownarrow$ \\
          & $J_{Ka, Kc} = 11_{4,8} - 11_{3,9}$   & 357.388 & $0.0054 \pm 0.0004$ & $1506.7 \pm 7.2$ & $133.1 \pm 4.7$ & $0.8 \pm 0.1$ \\
          & $J_{Ka, Kc} = 8_{4,4} - 8_{3,5}$     & 357.581 & $\downarrow$ \\
          & $J_{Ka, Kc} = 9_{4,6} - 9_{3,7}$     & 357.672 & $0.0122 \pm 0.0004$ & $1537.9 \pm 1.6$ & $98.6 \pm 3.8$ & $1.3 \pm 0.1$ \\
          & $J_{Ka, Kc} = 7_{4,4} - 7_{3,5}$     & 357.892 & $\downarrow$ \\
          & $J_{Ka, Kc} = 6_{4,2} - 6_{3,3}$     & 357.926 & $\downarrow$ \\
          & $J_{Ka, Kc} = 17_{4,14} - 17_{3,15}$ & 357.963 & $0.0290 \pm 0.0008$ & $1503.5 \pm 2.8$ & $165.1 \pm 3.9$ & $5.1 \pm 0.2$ \\
          & $J_{Ka, Kc} = 25_{3,23} - 25_{2,24}$ & 359.151 & $0.0118 \pm 0.0013$ & $1484.4 \pm 8.2$ & $157.7 \pm 19$ & $2.0 \pm 0.3$\\
H$_2$O    & $J_{Ka, Kc} = 5_{5,0} - 6_{4,3}, \ v_2 = 1 $ & 232.687 & $0.0043 \pm 0.0003$ & $1488.1 \pm 13$ & $344.7 \pm 33$ & $1.6 \pm 0.2$ \\
CN        & $N=3-2, \ J=5/2 - 3/2$      & 340.032 & $0.0105 \pm 0.0007$ & $1577.8 \pm 8.8$ & $239.9 \pm 7.8$ & $2.7 \pm 0.2$ \\
          & $N=3-2, \ J=7/2 - 5/2$      & 340.248 & $0.0230 \pm 0.0006$ & $1493.4 \pm 3.6$ & $239.9 \pm 7.8$ & $5.9 \pm 0.2$ \\
CS        & $J=7-6$                     & 342.883 & $0.0296 \pm 0.0009$ & $1501.2 \pm 3.4$ & $213.2 \pm 7.9$ & $6.7 \pm 0.3$ \\
HC$^{15}$N& $J=4-3, \ {\rm v}=0$              & 344.200 & $0.0081 \pm 0.0043$ & $1489.7 \pm 22$& $141.1 \pm 5.9$ & $1.2 \pm 0.6$ \\
H$^{13}$CN& $J=4-3, \ {\rm v}=0$              & 345.340 & $0.0594 \pm 0.0016$ & $1488.2 \pm 2.5$ & $187.4 \pm 5.8$ & $11.9 \pm 0.5$ \\
HCN       & $J=4-3, \ {\rm v}=0$              & 354.505 & $0.1508 \pm 0.0018$ & $1506.9 \pm 1.0$ & $168.3 \pm 2.4$ & $27.0 \pm 0.5$ \\
          & $J=4-3, \ {\rm v}_2=1f$           & 356.256 & $0.0803 \pm 0.0019$ & $1495.7 \pm 2.0$ & $152.6 \pm 2.9$ & $13.0 \pm 0.4$ \\
HCO$^+$   & $J=4-3, \ {\rm v}=0$              & 356.734 & $0.0980 \pm 0.0020$ & $1495.4 \pm 1.9$ & $152.6 \pm 2.9$ & $15.9 \pm 0.4$ \\
          & $J=4-3, \ {\rm v}_2=1e$           & 356.549 & $0.0274 \pm 0.0021$ & $1512.5 \pm 7.1$ & $152.6 \pm 2.9$ & $4.5 \pm 0.4$ \\
          & $J=4-3, \ {\rm v}_2=1f$           & 358.242 & $0.0303 \pm 0.0008$ & $1496.6 \pm 2.4$ & $165.1 \pm 3.9$ & $5.3 \pm 0.2$
\enddata
\tablecomments{ 
(1) Line species;  (2) Transition; (3) Rest frequency; (4) Peak optical depth; (5) LSR velocity at the peak optical depth; (6) Velocity width; (7) Equivalent width $= \int \tau dV$. SO$_2$ features of $J_{Ka, Kc} = 13_{4,10} - 13_{3,11}$, $15_{4,12} - 15_{3,13}$, and $11_{4,8} - 11_{3,9}$ are too close to decompose, thus combined parameters are listed. Combinations of ($8_{4,4} - 8_{3,5}$ and $9_{4,6} - 9_{3,7}$) and ($7_{4,4} - 7_{3,5}$, $6_{4,2} - 6_{3,3}$ and $17_{4,14} - 17_{3,15}$) are also indivisible, too.
}
\end{deluxetable} \label{tab:absorptionLineID}

The continuum spectrum of NGC 1052 casts a number of absorption features as shown in figures \ref{fig:B6Spectrum}, \ref{fig:B7LSBSpectrum}, and \ref{fig:B7USBSpectrum}.
CO ($J=2-1$ and $J=3-2$) absorption features clearly appear at the systemic velocity of 1492 km s$^{-1}$.
The velocity range\footnote{where absorbed flux density exceeds 3 times as large as the spectral image rms} of $J=2-1$ absorption spreads in 1224 km s$^{-1}$ -- 1773 km s$^{-1}$.
Although $J=3-2$ absorption feature involves contamination of SO $J_N = 3_2 - 1_2$ and H$^{13}$CN $J=4-3$ absorption features, the velocity range is consistent with $J=2-1$.
The velocity range covers CO ($J=1-0$) absorption observed with PdBI \citep{2004A&A...428..445L} which showed a shallower feature consisting of multiple velocity components. The difference will be discussed in \S \ref{subsec:CO}.

HCN ($J=4-3, \ v=0$) absorption appears around the systemic velocity, too, with the velocity range of 1174 km s$^{-1}$ -- 1874 km s$^{-1}$.
That covers multiple velocity components of HCN ($J=1-0, \ v=0$) absorption feature \citep{2004A&A...428..445L, 2016ApJ...830L...3S}, while $J=4-3$ feature appears in a single velocity component.
The difference will be discussed in \S \ref{subsec:HCN}.

HCO$^{+}$ ($J=4-3, \ v=0$) absorption is also clearly detected around the systemic velocity, which is consistent with the KVN observation of HCO$^{+}$ ($J=1-0$)\citep{2019ApJ...872L..21S}, but the velocity is offset from the PdBI observation \citep{2004A&A...428..445L}.
In addition to these known molecular species, we have identified CS, SO, SO$_2$, CN, and H$_2$O absorption lines.
Furthermore, isotopologues (H$^{13}$CN and HC$^{15}$N) and vibrationally excited (HCN v$_2 = 1$ and HCO$^{+}$ v$_2 = 1$) features are also identified.

Identified absorption lines are listed in table \ref{tab:absorptionLineID}, together with single Gaussian fitting to characterize the optical depth, LSR velocity, and line width as:
\begin{eqnarray}
S(v) = S_0 \exp \left(-\tau_0 \frac{(v - v_{\rm peak})^2}{2\sigma^2_v} \right), \label{eqn:GaussianTau}
\end{eqnarray}
where $S(v)$ is the observed flux density at the LSR velocity of $v$, $S_0$ is the continuum flux density, $\tau_0$ is the peak optical depth, $v_{\rm peak}$ is the LSR velocity at the peak optical depth, and $2 \sqrt{2 \ln 2} \sigma_v$ is the velocity width. 

For some line features that appear overlapped, we applied decomposition using multiple Gaussian components.
Identifications of these line species are discussed in \S \ref{subsec:lineID}.
No significant emission or absorption features with hydrogen recombination lines (H$26\alpha$: $\nu_{\rm rest} = 353.62275$ GHz, and H$30\alpha$: $\nu_{\rm rest} = 231.90093$ GHz) was detected.
We anticipated SiO $v=0$, $J=5-4$ ($\nu_{\rm rest} = 217.105$ GHz) and $J=8-7$ ($\nu_{\rm rest} = 347.331$ GHz) absorption features, but no significant features appear at the systemic velocity.
A bandpass remnant at 357.63 GHz is caused by the telluric O$_3$ $J_{Ka, Kc} = 20_{1,19} - 19_{2,18}$ attenuation.

\section{Discussion} \label{sec:discussion}

\subsection{Continuum emission} \label{subsec:continuumComponent}
Most of the unresolved continuum emission at the map center is dominated by the core, which is considered to be composed by the nucleus and compact jet components.
The size of the core is much smaller than the the synthesized beam of $0^{\prime \prime}.145$ in the direction of VLBI-scale jet.
The upper limit of the core size is $5.7 \times 5.1$ milliarcsec ($0.48 \times 0.43$ pc), based on a Gaussian fit in the continuum image.
The core component show a spectral index of $\alpha = -0.22 \pm 0.01$, which is consistent with the spectral index of the $\lambda$=3 mm VLBI measurements \citep{2016A&A...593A..47B, 2016ApJ...830L...3S, 2019ApJ...872L..21S}.

While VLA observations at 8.27 GHz detected kpc-scale jet and radio lobe components \citep{2020AJ....159...14N}, our ALMA observations do not show significant jet features with the upper limit of $0.035$ mJy beam$^{-1}$ ($3\sigma$ significance). Inferred spectral index of the jet is $\alpha < -1.0$, consistent with an optically thin synchrotron spectrum.

Although the dust absorption feature in the HST 1.6 $\mu$m image \citep{2001AJ....122..653R} coincides with the CND observed with CO emission (figure \ref{fig:Cont+CO}), no significant dust continuum emission is detected in our ALMA observations.
Because the minor axis of the CND is smaller than the maximum recoverable scale (see \S \ref{subsec:COemission}), the non-detection of dust emission is not due to missing flux.
Thus, the upper-limit ($3\sigma$) of dust brightness temperature in the CND is 0.28 K.

The component $6^{\prime \prime}$ to the north of the core is putatively a background sub-mm galaxy with the spectral index of $\alpha = 4.32 \pm 0.12$, which is consistent with dust emission spectrum.

\begin{figure}[ht]
\begin{center}
\includegraphics[scale=0.51,angle=0]{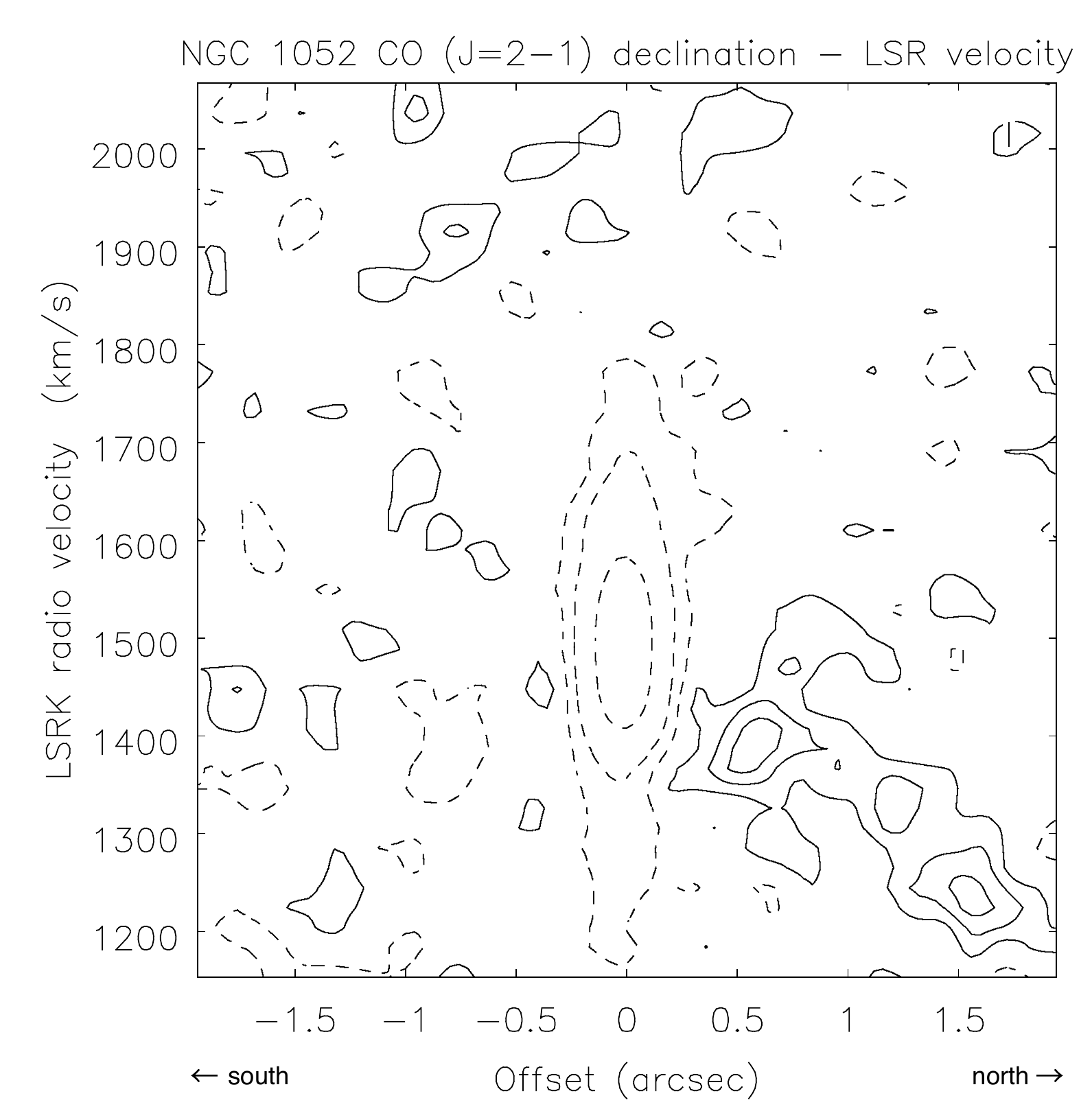}
\includegraphics[scale=0.53,angle=0]{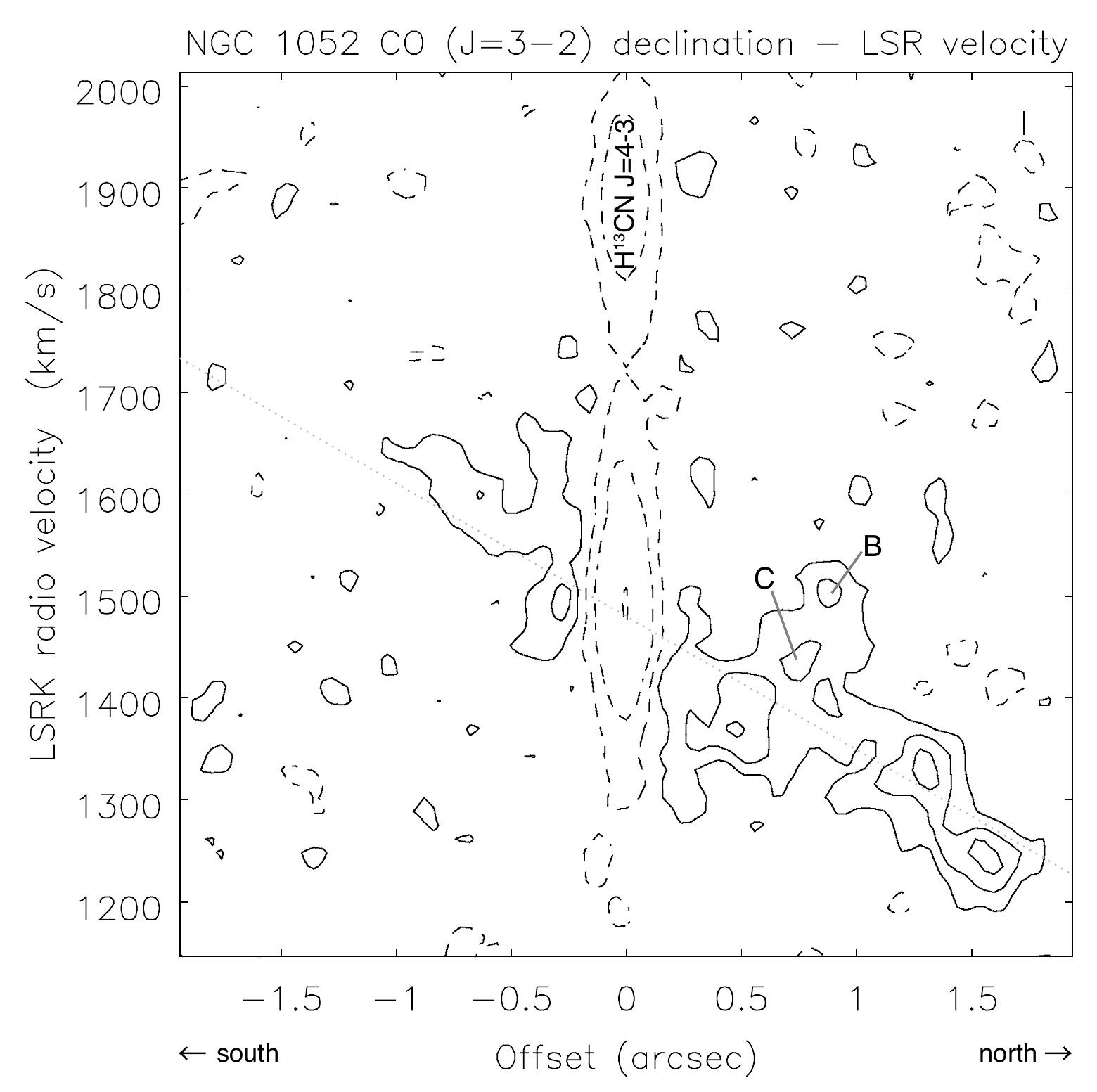}
\caption{Position--velocity diagrams of CO $J=2-1$ (left) and $J=3-2$ (right) across the core in north--south direction (dashed-line rectangles in figure \ref{fig:COvelocity}). The negative component at the center indicates the CO absorption line. The blobs marked B and C stand for the velocity components in table \ref{tab:emissionDecomposition}. The dotted diagonal line on the right panel indicates linear regression with the systemic velocity of $1492.0 \pm 2.6$ km s$^{-1}$ and the velocity gradient of $136.3 \pm 2.7$ km s$^{-1}$ arcsec$^{-1}$. \label{fig:COPV}}
\end{center}
\end{figure}

\begin{figure}[ht]
\begin{center}
\includegraphics[scale=0.8,angle=0]{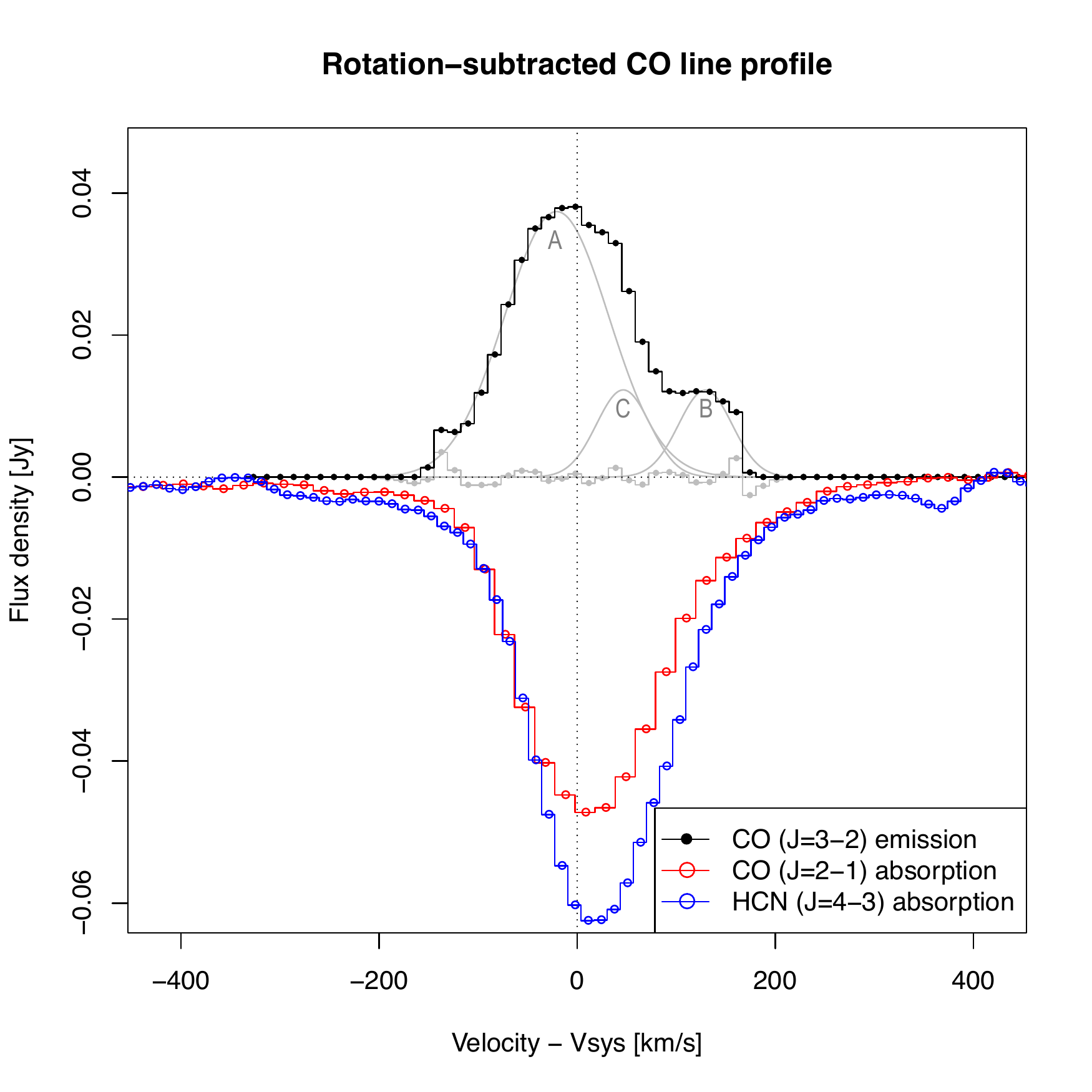}
\caption{Black solid line: CO ($J=3-2$) emission line profile (black) integrated in the position--velocity diagram of figure \ref{fig:COPV}, clipped by the threshold of 0.1 mJy/beam, and de-rotation by the principal component of the velocity gradient of $136.3 \pm 2.7$ km s$^{-1}$ arcsec$^{-1}$ and the intercept at $1492.0 \pm 2.6$ km s$^{-1}$. The de-rotation emission profile consists of three Gaussian components and the residual shown in grey lines. Parameters of the Gaussian components are listed in table \ref{tab:emissionDecomposition}.
Red and blue solid lines: absorption features of CO $J=2-1$ and HCN $J=4-3$, respectively, toward the continuum core. CO $J=3-2$ absorption profile is not presented here because of H$^{13}$CN absorption feature overlapped (see figure \ref{fig:B7LSBSpectrum}).}
\label{fig:rotationSubtract}
\end{center}
\end{figure}

\begin{deluxetable}{lrrr}
\tablecaption{Gaussian decomposition of rotation-subtracted CO ($J=3-2$) emission profile}
\tablehead{
\colhead{Component} &  \colhead{$S_{\rm peak}$}  &  \colhead{$V_{\rm peak} - V_{\rm sys}$} & \colhead{FWHM} \\
\colhead{} & \colhead{(mJy)} &  \colhead{(km s$^{-1}$)}  &  \colhead{(km s$^{-1}$)}  
}
 \colnumbers
 \startdata
A  & $37.4 \pm 1.2$ & $-20.7 \pm 4.8$ & $123.7 \pm 6.5 $ \\
B  & $12.3 \pm 0.6$ & $129.5 \pm 2.3$ & $ 63.6 \pm 5.3 $ \\
C  & $12.3 \pm 3.7$ & $ 46.5 \pm 2.9$ & $ 63.7 \pm 11 $ \\
\enddata
\tablecomments{ 
(1) Velocity component;  (2) Peak flux density (mJy); (3) Peak velocity with respective to the systemic velocity after subtraction of rotation (km s$^{-1}$); (4) Velocity width (FWHM, km s$^{-1}$).
}
\end{deluxetable} \label{tab:emissionDecomposition}

\subsection{Circum-Nuclear Disk} \label{subsec:CND}
The identified CND extends 230 pc and 60 pc in the north-south and east-west directions, respectively.
Orientations of the CND major axis\footnote{Estimated by $\frac{1}{2}$ atan2$(2M_{1,1},  M_{2,0} - M_{0,2})$ where $M_{i, j}$ stands for $i$-th moment of $l$ and $j$-th moment of $m$. Velocity gradient is calculated using velocity-weighted moments.} is $5^{\circ}.0 \pm 1^{\circ}.7$ while that of velocity gradient is $4^{\circ}.5 \pm 1^{\circ}.7$.
Hereafter, we employ the north-south direction as the major axis for simplicity.
The velocity field indicates that the CND consists of a ring structure, rather than a filled disk, seen edge-on.
Figure \ref{fig:COPV} shows the position--velocity diagram of CO $J=2-1$ and $J=3-2$ emission.
The aperture for the diagram is indicated by a dashed-line rectangle that covers the CND.
The diagram follows the diagonal line, determined by a linear regression\footnote{Solving for $a$ and $b$ to minimize $\int \int I(m, v) (v - am - b)^2 \ dm \ dv $, where $I(m, v)$ is the intensity at the offset, $m$, and the velocity, $v$. $a$ and $b$ stands for the velocity gradient and the systemic velocity, respectively.} on the position-velocity diagram of CO $J=3-2$ emission shown in the dotted line in figure \ref{fig:COPV}.

The diagram follows the diagonal line with the velocity gradient of $136.3 \pm 2.7$ km s$^{-1}$ arcsec$^{-1}$ with the intercept at $1492.0 \pm 2.6$ km s$^{-1}$.
A rotating molecular ring, seen edge-on, accounts for this principle component.
If we assume the CND is rotationally supported without significant radial motions, the intercept indicates the systemic velocity of the dynamical center.

The rotation curve, with the terminal velocity of $|V_{\rm LSR} - V_{\rm sys}|_{\rm max} = 245 \pm 4.9$ km s$^{-1}$ at the radius of 153 pc suggests the enclosed mass, $M_{\rm enc}$, inside the CND as, $\displaystyle M_{\rm enc} = \frac{RV^2}{G} = (2.13 \pm 0.09) \times 10^9$ M$_{\odot}$.
That is significantly greater than the black-hole mass of $1.5 \times 10^8$ M$_{\odot}$ derived with stellar velocity dispersion \citep{2002ApJ...579..530W}.
Therefore, the CND resides outside of the black-hole sphere of influence (SoI) with a radius of 14.4 pc estimated in \S \ref{subsec:angularMomentum}.


While most of CO emission is ascribed to the principal component of the rotating ring, we identify extra emission-line components in the P--V diagram.
Figure \ref{fig:rotationSubtract} shows the rotation-subtracted CO emission profile\footnote{derived by $\int I(m, v - am - b) dm$.} which consists of three Gaussian velocity components listed in table \ref{tab:emissionDecomposition}.
Component A stands for the rotating ring, while B and C reside at $\Delta \delta = 0^{\prime \prime}.8$ and $0^{\prime \prime}.7$ in the P--V diagram, respectively, as marked in figure \ref{fig:COPV}.
The velocity width of component A is significantly wider than that of B and C.
If the excess in the velocity width, $\Delta V = \sqrt{\sigma^2_{V(A)} - \sigma^2_{V(B)}} = 45$ km s$^{-1}$, is caused by differential Keplerian rotation, the radial width would be $\Delta R = 2 GM_{\rm enc} (V_{\rm LSR} - V_{\rm sys})^{-3} \Delta V = 56$ pc.
That is almost the same with the E-W extent of the CND which would be the thickness of the CND if seen edge-on.
If we employ the mean diameter of 230 pc and the mean width of 58 pc for the ring-like CND, its volume is estimated to be $2\times 10^6$ pc$^3$.

No high-velocity emission is identified inside the CND.
This indicates that CND structure is ring-like with a void, unlike a filled disk as studied by \citet{2016ApJ...823...51B} for NGC 1332.
The void is not an artifact caused by bright continuum emission with the strong absorption feature but a real one --- our synthesized beam is sufficiently sharp to resolve the CND from the core.
We have no clue why CO emission is absent inside the CND ring. As we discuss in the next subsection, there must be a pc-scale massive molecular torus much closer to the core.
The connection between the CND and the torus is apparently missing in NGC 1052.

The edge-on structure of the CND is asymmetric; the northern part is significantly brighter with a greater extent than the southern part.
The CO emission peak in the northern end of the CND seems morphologically connected to outer arc structure (see $V_{\rm LSR} = 1205 - 1340$ km s$^{-1}$ channels in figures \ref{fig:CO21ChanMap} and \ref{fig:CO32ChanMap}).
Such an asymmetric CND structure is similar to that in NGC 1068 \citep{2014A&A...567A.125G, 2018ApJ...853L..25I}, where CO emission peak is connected to the outer starburst ring by a bridge of molecular material (labeled the arc).
The arc would be a gas supplier into the CND. The asymmetry might be caused by sporadic infall of molecular clouds into the CND, which is not dynamically relaxed yet with the orbital period of $2\times 10^6$ yr.

We estimated the molecular gas mass in the CND as $M_{\rm H_2} = 5.3\times 10^5$ M$_{\odot}$, based on the CO luminosities of $L^{\prime}_{\rm CO \ 2-1} = (2.89 \pm 0.26) \times 10^5$ K km s$^{-1}$ pc$^2$ and $L^{\prime}_{\rm CO \ 3-2} = (6.58 \pm 0.28) \times 10^5$ K km s$^{-1}$ pc$^2$ for $J=2-1$ and $J=3-2$ transitions, respectively (see table \ref{tab:MCflux}).
The greater luminosity at the higher-level transition indicates that CO is thermalized with a high excitation temperature of $T_{\rm ex} \gg hBJ(J+1)/k = 33$ K.
The luminosity in $J=2-1$ could be underestimated because of confusion of absorption features with the twice wider solid angle of the synthesized beam at lower frequency, because other molecular clouds apart from the core show $L^{\prime}_{\rm CO \ 2-1} \simeq L^{\prime}_{\rm CO \ 3-2}$. 
Applying the method to derive molecular mass from $L^{\prime}_{\rm CO \ 3-2}$ \citep{2018ApJ...867...48I} with the CO conversion factor of $\alpha_{\rm CO (1-0)} = 0.8$ M$_{\odot}$ (K km s$^{-1}$ pc$^2$)$^{-1}$ and the thermalized condition for CO $J=1-0$ and $J=3-2$ transitions, we obtain the molecular gas mass above.
This CO conversion factor would be too small for a non-starburst galaxy like NGC 1052 then the molecular mass would be underestimated by a factor of $\sim 10$.
If we apply the estimation for CND gas mass in Cen A with the conversion factor between the integrated CO intensity and H$_2$ column density, $X_{\rm CO} = 5 \times 10^{20}$ cm$^{-2}$ K$^{-1}$  km$^{-1}$ s, and the flux ratio of $S_{\rm CO (1-0)} / S_{\rm CO (3-2)}  \simeq 0.1$ \citep{2017ApJ...843..136E, 2019ApJ...887...88E}, we have $\alpha_{\rm CO (1-0)} = 7.3$ M$_{\odot}$ (K km s$^{-1}$ pc$^2$)$^{-1}$ and $M_{\rm H_2} = 4.8\times 10^6$ M$_{\odot}$.
Nevertheless, the estimated CND gas mass is smaller than typical dense molecular gas mass in Seyfert galaxies of $10^{7-8}$ M$_{\odot}$ \citep{2016ApJ...827...81I} or comparable to the Circinus galaxy: $M_{\rm H_2} \sim 3 \times 10^6$ M$_{\odot}$ \citep{2018ApJ...867...48I}.

The molecular density, $n_{\rm H_2}$, must exceed the critical density of $4 \times 10^4$ cm$^{-3}$ for CO $J=3-2$ emission if radiative trapping is negligible.
Absence of HCN and HCO$^+$ $J=4-3$ emissions indicates that the molecular density is lower than the critical densities of $3.2 \times 10^7$ cm$^{-3}$ and $5.6 \times 10^7$ cm$^{-3}$, respectively.
The mean molecular density in the CND of $n_{\rm H_2} \sim 6 \times 10^6$ cm$^{-3}$, estimated by dividing the estimated gas mass by the volume of the ring-like CND, is consistent with the conditions of critical densities for CO, HCN, and HCO$^+$.

\cite{2016ApJ...827...81I} found a tight correlation between CND-scale dense gas mass and mass accretion rate in 10 Seyfert galaxies and proposed that star formation in CNDs drives mass accretion onto SMBHs. This scenario does not work for NGC 1052 which does not align with the correlation. $L_{\rm bol} = 6.92 \times 10^{43}$ erg s$^{-1}$ \citep{2002ApJ...579..530W} yields an accretion rate of $\dot{M}_{\rm BH} = 0.01$ M$_{\odot}$ yr$^{-1}$ if we adopt a mass-to-energy conversion factor of 0.1.
For that accretion rate, the correlation predicts a gas mass of $1.3 \times 10^8$ M$_{\odot}$, which is far from our measurement by two or more orders of magnitude.

The Toomre's $\displaystyle Q = \frac{\sigma_R \kappa}{3.36 G \Sigma}$ is estimated to be $\sim 500$, using the surface density of $\displaystyle \Sigma = \frac{M_{\rm H_2}}{2 \pi R_{\rm mean} \Delta R} = 13$ M$_{\odot}$ pc$^{-2}$, the radial velocity width of $\sigma_R = 63.6$ km s$^{-1}$, and the epicyclic frequency, $\kappa = \sqrt{2} V /R$, for a flat rotation curve.
Thus we conclude that the CND in NGC 1052 is too gas-poor to drive mass accretion via ongoing star formation.
We cannot identify any CO molecular clouds or the CND associated with the young stellar clusters discovered by near infrared observations \citep{2011MNRAS.411L..21F}.

\subsection{Absorption line identification} \label{subsec:lineID}
This subsection addresses absorption line features in the continuum spectrum of the core component shown in figures \ref{fig:B6Spectrum}, \ref{fig:B7LSBSpectrum}, and \ref{fig:B7USBSpectrum}.
The spectrum of a single pixel at the map center covers whole continuum flux of the core with absorption caused by gas in front of the core.
Identified absorption features are listed in table \ref{tab:absorptionLineID}, together with remarks on characterization for each line feature by Gaussian fitting.

\subsubsection{CO} \label{subsec:CO}
Carbon monoxide absorption features of $J=2-1$ ($\nu_{\rm rest} = 230.5380$ GHz) and $J=3-2$ ($\nu_{\rm rest} = 345.796$ GHz) clearly appear at the peak LSR velocity of $1496.7 \pm 1.4$ km s$^{-1}$ and $1494.7 \pm 1.1$ km s$^{-1}$. The peak optical depths are $0.0993 \pm 0.0017$ and $0.1284 \pm 0.0017$, respectively.
The line profile of $J=2-1$ extends between 1160 -- 1900 km s$^{-1}$ with a skewed wing in the redshifted side.
Because H$^{13}$CN absorption feature overlaps the outskirt of CO $J=3-2$, we decomposed them by fitting with two Gaussian components. We omit contamination of SO $J_N = 3_2 - 1_2$ because the line strength ($-5.8873$ in CDMS/JPL intensity) is weaker than that of H$^{13}$CN ($-0.7877$).

Our line profiles of $J=2-1$ and $J=3-2$ with ALMA are significantly different from $J=1-0$ absorption profile with the PdBI \citep{2004A&A...428..445L}, where the optical depth peaks around $\sim 1700$ km s$^{-1}$ and the absorption is absent around the systemic velocity.
There are two possible reasons to explain the difference. First, the beam size of the PdBI\footnote{The beam size was not explicitly described in \citet{2004A&A...428..445L}. According to the PdBI observation log, they employed D configuration that offers a $\sim 6^{\prime \prime}$ resolution at 100 GHz.} was $\sim 6^{\prime \prime}$ that must have included CO emission in the CND, while our ALMA observation has a $0^{\prime \prime}.209 \times 0^{\prime \prime}.145$ beam that is less contaminated with the emission. Second, the continuum structure at 115 GHz is supposed to be larger than that at 344 GHz due to steep spectrum of jet components, and thus the covering factor of molecular clumps would be smaller at lower frequencies to show shallower absorption.

\subsubsection{HCN} \label{subsec:HCN}
Hydrogen cyanide H$^{12}$C$^{14}$N $J=4-3, \ {\rm v}=0$ ($\nu_{\rm rest} = 354.505$ GHz) is the deepest absorption feature in our observations with the peak optical depth of $0.1508 \pm 0.0018$ at $1506.9 \pm 1.0$ km s$^{-1}$.
The line profile is consistent with that in CO $J=2-1$, with a skewed skirt, as shown in figure \ref{fig:rotationSubtract}.

The absorption profile is significantly different from that in $J=1-0$ observed with PdBI \citep{2004A&A...428..445L}, as well as the difference in CO absorption profile. While HCN $J=4-3$ emission was not detected in the CND, $J=1-0$ emission might cause contamination in the absorption profile of PdBI because the estimated molecular density in \S \ref{subsec:CND} exceeds the critical density of HCN  $J=1-0$ transition.

In addition to H$^{12}$C$^{14}$N, we identified isotopologues of H$^{13}$CN ($\nu_{\rm rest} = 345.340$ GHz) and HC$^{15}$N ($\nu_{\rm rest} = 344.200$ GHz) absorption features. We applied two-Gaussian decomposition to discriminate H$^{13}$CN from CO $J=3-2$. 
Because HC$^{15}$N appeared close to SO $J_N = 8_8 - 7_7$ ($\nu_{\rm rest} = 344.3106$ GHz), the decomposition results in relatively large uncertainty in the optical depth of $\tau_{\rm max} = 0.0081 \pm 0.0043$.

We also identified vib-excited H$^{12}$C$^{14}$N $J=4-3, \ {\rm v}_2=1, \ \ell=1f$ ($\nu_{\rm rest} = 356.256$ GHz) feature near the HCO$^+$ features.
Another vib-excited line of ${\rm v}_2=1, \ \ell=1e$ ($\nu_{\rm rest} = 356.460$ GHz) is expected to have a similar optical depth of $\ell=1f$ \citep{2016A&A...590A..25M}. However, the frequency offset from the ground-state line is only 45 MHz (38 km s$^{-1}$), which cannot be decomposed. Since the HCN ground-state line is considered to be optically thick (see \S \ref{subsec:coveringFactor}), we do not take the contribution of ${\rm v}_2=1, \ \ell=1e$ into account.

\subsubsection{HCO$^{+}$}
Formylium $J=4-3, \ v=0$ ($\nu_{\rm rest} =356.734$ GHz) show the peak optical depth of $0.0980 \pm 0.0020$ at $1495.4 \pm 1.9$ km s$^{-1}$.
Near the HCO$^{+}$ main feature, we identified vibrationally excited HCO$^+$ $J=4-3, \ {\rm v}_2=1, \ \ell=1e$ ($\nu_{\rm rest} = 356.549$ GHz) feature. We applied two-Gaussian decomposition to characterize these features with a common line width.
Relatively large uncertainty in the LSR velocity of the vib-excited feature is caused by proximity to the $v=0$ line.
Another vibrationally excited line of $J=4-3, \ {\rm v}_2=1, \ \ell=1f$ ($\nu_{\rm rest} = 358.242$ GHz) is identified, which is clearly isolated from adjacent features of SO$_2$.

\subsubsection{SO}
Sulfur monoxide features appear in transitions of $J_N=7_8 - 7_7$ ($\nu_{\rm rest} = 214.357$ GHz), $J_N=5_5 - 4_4$ ($\nu_{\rm rest} = 215.220$ GHz), $J_N=7_8 - 6_7$ ($\nu_{\rm rest} = 340.714$ GHz), $J_N=8_8 - 7_7$ ($\nu_{\rm rest} = 344.311$ GHz), and $J_N=9_8 - 8_7$ ($\nu_{\rm rest} = 346.528$ GHz).
The feature at $\nu_{\rm topo} = 214.8$ GHz is ascribed to the isotopologue $^{34}$SO $J_N=6_5 - 5_4$ ($\nu_{\rm rest} = 215.839$ GHz), with a possible contamination of CH$_3$OH $J_{Ka, Kc} = 8_{4,5} - 9_{3,6}$ (215.708 GHz) and $J_{Ka, Kc} = 4_{2,2} - 3_{1,2}$ (215.887 GHz).
These absorption features are not significantly affected by line overlaps. The $J_N=7_8 - 7_7$ absorption is close to HC$^{15}$N $J=4-3$ feature but its contamination is minor.
The $J_N = 8_9 - 7_8$ feature could be contaminated by SO$_2$ $J_{Ka, Kc} = 16_{4,12} - 16_{3,13}$ transition at $\nu_{\rm rest} = 346.524$ GHz, which is too close to be decomposed. The differences in LSR velocity and FWHM of the $J_N = 8_9 - 7_8$ feature from other transitions can be caused by the SO$_2$ line contamination.

\subsubsection{SO$_2$}
There are 14 line features that correspond to SO$_2$ transitions appeared around the systemic velocity, with $J$ ranging 6 -- 40 (41.4 K $\leq E_L \leq 792.0$ K).
Some groups of transitions reside close to each other in the spectrum: $J_{Ka, Kc} = 21_{8,14} - 22_{7,15}, \ 40_{4,36} - 40_{3, 37}, \ 36_{5, 31} - 36_{4,32}$ around $\nu_{\rm topo} \sim 339.7$ GHz,
$J_{Ka, Kc} = 13_{4,10} - 13_{3,11}, \ 15_{4,12} - 15_{3, 13}, \ 11_{4, 8} - 11_{3,9}$ around $\nu_{\rm topo} \sim 355.5$ GHz, 
$J_{Ka, Kc} = 8_{4,4} - 8_{3,5}, \ 9_{4,6} - 9_{3, 7}$ around $\nu_{\rm topo} \sim 355.8$ GHz,
and $J_{Ka, Kc} = 7_{4,4} - 7_{3,5}, \ 6_{4,2} - 6_{3, 3}, \ 17_{4, 14} - 17_{3,15}$ around $\nu_{\rm topo} \sim 356.1$ GHz.
We applied three-component Gaussian decomposition for the 339.7-GHz group with a common velocity width.
For the 355.5-GHz, three-Gaussian decomposition does not converge. Then, we applied two-Gaussian decomposition where $13_{4,10} - 13_{3,11}$ and $15_{4,12} - 15_{3,13}$ features are unresolved.
The 355.8-GHz and group is totally unresolved where only one-Gaussian decomposition works for a mixture of $8_{4,4} - 8_{3,5}$ and $9_{4,6} - 9_{3,7}$.
Three transitions, $7_{4,4} - 7_{3,5}$, $6_{4,2} - 6_{3,3}$, and $17_{4,14} - 17_{3,15}$ in the 356.1-GHz group are also unresolved.

\subsubsection{CS}
Carbon monosulfide $J=7-6$ ($\nu_{\rm rest} = 342.883$ GHz) feature clearly appears, with a plumper profile compared with HCN $J=4-3, \ v=0$. SO$_2$ $J_{Ka, Kc} = 34_{3,31} - 34_{2,32}$ ($\nu_{\rm rest} = 342.762$ GHz) and H$_2$CS $J_{Ka, Kc} = 10_{0,10} - 9_{0,9}$ ($\nu_{\rm rest} = 342.944$ GHz) might contribute to the profile.

\subsubsection{CN}
Cyanide radical transitions of $N=3-2, \ J=5/2 - 3/2$ with 6 hyperfine structures and $J=7/2 - 5/2$ with 6 hyperfine structures appear around $\nu_{\rm topo} \sim 338 - 339$ GHz. The hyperfine structures, too close to be resolved, form a blended absorption profile.

\subsubsection{H$_2$O}
There is a shallow absorption feature around $\nu_{\rm topo} \sim 231.5$ GHz. The frequency slightly differs from the telluric O$_3$ $J_{Ka, Kc} = 16_{1,15} - 16_{0,16}$ absorption feature at 231.281511 GHz.
The feature could be identified as H$_2$O $J_{Ka, Kc} = 5_{5, 0} - 6_{4, 3}, \ v_2=1$ absorption in NGC 1052.
The line profile is not as simple as other molecular absorption features such as HCN $J=4-3, \ v=0$, rather resembles the H$_2$O $J_{Ka, Kc} = 6_{1, 6} - 5_{2, 3}$ maser emission at 22 GHz that spreads in 1450 km s$^{-1} \leq V_{\rm LSR} \leq 1850$ km s$^{-1}$ \citep{2003ApJS..146..249B, 2005ApJ...620..145K}.
The $v_2=1$ state requires the energy level of 3451 K, which is greater than the condition of $T\sim 10^3$ K in the X-ray dissociation region (XDR) where H$_2$O maser is excitated \citep{2005ApJ...620..145K}.

\subsubsection{Absence of SiO}
Silicon monoxide transitions of $J=5-4, \ v=0$ ($\nu_{\rm rest} = 217.105$ GHz) and $J=8-7, \ v=0$ ($\nu_{\rm rest} = 347.331$ GHz) do not significantly appear in the spectrum, with the upper limit of $\tau_{\rm max} < 0.003$, while the frequency range covers them.

\subsection{Column densities of absorption lines} \label{subsec:columnDensity}
We applied single-Gaussian fitting for CO, HCN,  HCO$^+$, SO and CS to characterize the absorption features in terms of velocity widths and optical depths as listed in table \ref{tab:absorption}.

The total molecular column density, $N_{\rm tot}$, of each molecular species is derived assuming local thermodynamic equilibrium (LTE) as 

\begin{equation}
N_{\rm tot} = \frac{3 k T_{\rm ex}}{8 \pi^3 \mu^2 B (2J+1)} \exp{\left(\frac{hB}{3kT_{\rm ex}}\right)} 
\exp{\left( \frac{hBJ(J+1)}{kT_{\rm ex}} \right)} 
\left[ \exp{\left( \frac{h\nu}{kT_{\rm ex}}\right)} -1 \right]^{-1}
\int \tau dV, \label{eqn:columnDensity}
\end{equation}
where $k$ is the Boltzmann constant, 
$h$ is the Planck constant, 
$\mu$ is the permanent dipole moment of the molecule, 
$B$ is the rotational constant, $T_{ex}$ is the excitation temperature, 
$\nu$ is the rest frequencies, 
and $\int \tau dV$ is the equivalent width of the absorption feature.

Note that abundant line species such as CO, HCN, and HCO$^{+}$ are supposed to be optically thick with a small covering factor and the column densities estimated by equation \ref{eqn:columnDensity} is likely to be underestimated. We will discuss about correction for the optical depths and column densities in section \ref{subsec:coveringFactor}.

\begin{deluxetable}{lrrrrr}
\tablecaption{Column densities}
\tablehead{
\colhead{Line} & \colhead{$\nu_{\rm rest}$} & \colhead{$\mu$}   & \colhead{$B$}   & \colhead{$N_{\rm tot}$} & \colhead{$N_{\rm cor}$}  \\
\colhead{}     & \colhead{(MHz)}            & \colhead{(Debye)} & \colhead{(MHz)} & \colhead{(cm$^{-2}$)}   & \colhead{(cm$^{-2}$)}
 }
 \colnumbers
 \startdata
CO $J=2-1$ 	                     & 230538 & 0.1101 & 57635.96 & $(6.73 \pm 0.19) \times 10^{18}$ \\
CO $J=3-2$ 	                     & 345790 & 0.1101 & 57635.96 & $(3.66 \pm 0.08) \times 10^{18}$ & $(1.5 \pm 0.2) \times 10^{20}$ \\ 
HCN $J=4-3, \ v=0$               & 354505 & 2.984  & 44315.97 & $(5.48 \pm 0.10) \times 10^{15}$ & $(2.5 \pm 0.6) \times 10^{17}$ \\ 
HCN $J=4-3, \ {\rm v}_2=1f$            & 356256 & 2.942  & 44422.42 & $(2.69 \pm 0.08) \times 10^{15}$ \\
H$^{13}$CN $J=4-3, \ {\rm v}=0$        & 345340 & 2.985  & 43170.13 & $(2.54 \pm 0.11) \times 10^{15}$ & $(5.1 \pm 1.2) \times 10^{15}$ \\ 
HC$^{15}$N $J=4-3, \ {\rm v}=0$        & 344200 & 2.985  & 43027.64 & $(2.58 \pm 1.3) \times 10^{14}$ \\
HCO$^+$ $J=4-3, \ {\rm v}=0$           & 356734 & 3.888  & 44594.40  & $(1.88 \pm 0.05) \times 10^{15}$ & $(7.8 \pm 0.8) \times 10^{16}$ \\
HCO$^+$ $J=4-3, \ {\rm v}_2=1e$        & 356549 & 3.900 & 44677.15 & $(5.28 \pm 0.47) \times 10^{14}$ \\
HCO$^+$ $J=4-3, \ {\rm v}_2=1f$        & 358242 & 3.900 & 44677.15 & $(6.18 \pm 0.23) \times 10^{14}$ \\
SO $J_N = 8_7 - 7_7$             & 214357 & 1.535  & 21523.56 & $(1.28 \pm 0.14) \times 10^{15}$ \\
SO $J_N = 5_5 - 4_4$             & 215221 & 1.535  & 21523.56 & $(9.43 \pm 0.20) \times 10^{15}$ \\
SO $J_N = 8_7 - 7_6$             & 340714 & 1.535  & 21523.56 & $(7.15 \pm 0.44) \times 10^{15}$ \\
SO $J_N = 8_8 - 7_7$             & 344311 & 1.535  & 21523.56 & $(6.90 \pm 0.44) \times 10^{15}$ \\
SO $J_N = 8_9 - 7_8$             & 346528 & 1.535  & 21523.56 & $(1.01 \pm 0.05) \times 10^{16}$ \\
$^{34}$SO $J_N = 5_6 - 4_5$      & 215839 & 1.535  & 21102.73  & $(1.83 \pm 0.20) \times 10^{15}$ \\
CS $J = 7 - 6$                   & 342883 & 1.957  & 24495.56  & $(3.53 \pm 0.16) \times 10^{15}$
\enddata
\tablecomments{ 
(1) Line species; (2) Rest frequencies; (2) Dipole moments; (3) Rotational constants; (6) Total molecular column densities assuming $T_{\rm ex}=230 $ K; (7) Column densities of abundant molecular species after optical depth correction discussed in \S \ref{subsec:coveringFactor}.
}
\end{deluxetable} \label{tab:absorption}


\subsection{Molecular Torus} \label{subsec:TORUS}
We consider that the absorption features listed in \S \ref{subsec:lineID} are ascribed to a molecular torus surrounding the core rather than the CND.
While CO is the only emission line in the CND, several molecular species exhibits absorption features.
This indicates that the absorber resides in different physical environment from the CND.
The CND cannot account for the velocity width of absorption features, shown in figure \ref{fig:COPV}.
The velocity width of CO absorption features are FWHM $= 166.5 \pm 3.4$ km s$^{-1}$ and $=168.8 \pm 2.8$ km s$^{-1}$, which are significantly wider than that in CND of FWHM $= 123.7 \pm 6.5$ km s$^{-1}$.
The most prominent absorption line, HCN ($J=4-3$, v$=0$), shows the significant absorption feature in at least $V_{\rm LSR} = 1174 - 1874$ km s$^{-1}$, consisting of a Gaussian component with a wide and shallow wing.
The wing velocity width of $\Delta V_{\rm wing} = ^{+382}_{-318} $ km s$^{-1}$ exceeds the CND terminal velocity.

If we assume that the virial mass inferred from the wing corresponds to $M_{\rm BH} = 1.54 \times 10^8$ M$_{\odot}$ \citep{2002ApJ...579..530W}, the virial radius will be
\begin{eqnarray}
r_{\rm vir} = GM_{\rm BH} / \Delta V^2_{\rm wing} = 7.4 \ {\rm pc} \label{eqn:r_vir}
\end{eqnarray}
Note that $r_{\rm vir}$ gives us the upper limit of the torus radius because the continuum source is more compact than $5.7 \times 5.1$ milliarcsec (see \S \ref{subsec:continuumComponent}), rotation of CND with the velocity gradient of $136.3 \pm 2.7$ km s$^{-1}$ arcsec$^{-1}$ contributes less than 0.78 km s$^{-1}$ to the absorption line width.
The upper limit of the radius is significantly smaller than the CND size, and comparable to the molecular torus in NGC 1068; 12 pc $\times 7$ pc in CO ($J=6-5$) \citep{2016ApJ...829L...7G}, 13 pc $\times 4$ pc in HCN, and 12 pc $\times 5$ pc in HCO$^{+}$ \citep{2018ApJ...853L..25I} or $28 \pm 0.6$ pc in CO ($J=2-1$), $26 \pm 0.6$ pc in CO ($J=3-2$), $11 \pm 0.6$ pc in HCO$^+$ ($J=4-3$) \citep{2019A&A...632A..61G}, or inner and outer radii of 0.5 -- 1.2 and 7 pc in HCN ($J=3-2$), respectively \citep{2019ApJ...884L..28I}.

\begin{figure}[ht]
\begin{center}
\includegraphics[scale=0.6,angle=0]{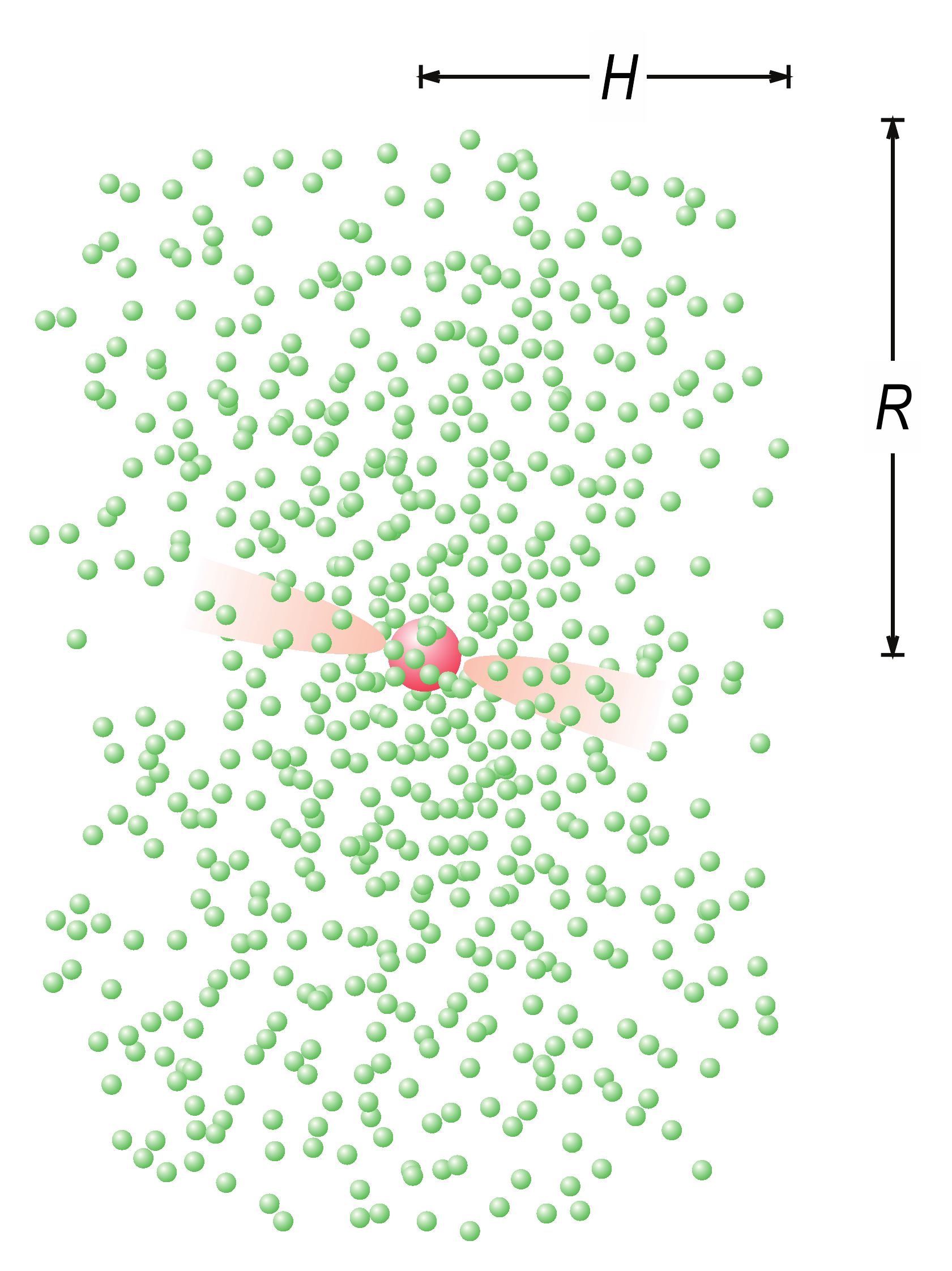}
\caption{A schematic diagram of a molecular torus. The torus is composed by molecular clumps (green blobs) surrounding the continuum radio core (red sphere) and jets (vermilion cones). The torus has a radius of $R$ and the height of $H$, and is seen edge-on.}
\label{fig:clumpyTorus}
\end{center}
\end{figure}

We here attempt estimating the height, $H$, the radius, $R$, and the rotation velocity, $V_{\rm rot}$, assuming a torus model illustrated in figure \ref{fig:clumpyTorus}. Let $\Theta = H/R$ the geometrical thickness of the torus, $V_H$ the non-rotational random velocity component, then we have $\Theta^2 = \left< V^2_H \right> / \left< V^2_{\rm rot} \right>$ for a torus in hydrostatic equilibrium\footnote{We do not take outflow into account because the bipolar outflow observed with [O III] \citep{2005ApJ...629..131S} is collimated in the intrinsic maximum opening angle of $2\alpha \sim 53^{\circ}$ at $1^{\prime \prime} - 1^{\prime \prime}.3$ away from the nucleus with the viewing angle of $57^{\circ} - 72^{\circ}$. Thus the bipolar outflow does not cover the compact radio continuum component.}.
The kinetic velocity dispersion will be $\left< V^2 \right> = \left< V^2_{\rm rot} \right> + \left< V^2_H \right> = \left< V^2_H \right> (1 + \Theta^{-2})$.
Because the countinuum source is compact, contribution of rotation velocity to the absorption width is 60 km s$^{-1}$ at most (see \S \ref{subsec:angularMomentum}) and $\left< V^2_H \right>$ dominates the velocity width of the absorption feature, $\Delta V^2_{\rm wing}$.
The virial radius estimated in equation \ref{eqn:r_vir} should be corrected as
\begin{eqnarray}
R = \frac{GM_{\rm BH}}{(1 + \Theta^{-2}) \Delta V^2_{\rm wing}} = \frac{r_{\rm vir}}{1 + \Theta^{-2}}. \label{eqn:virialRadius}
\end{eqnarray}
If we assume that the torus radius is greater than the radius of plasma torus which resides in $\sim 1$ pc \citep{2001PASJ...53..169K, 2003PASA...20..134K}, $R \geq 1$ pc gives $\Theta \geq 0.4$ and $V_{\rm rot} < 750$ km s$^{-1}$.
The H$_2$O maser distribution in 0.2 pc (projected distance from the core along the jet) \citep{2008ApJ...680..191S} yields $H \geq 0.2$ pc and thus $\Theta \geq 0.3$, consistently.

Further consideration on the physical parameters of the torus is followed by the next subsection.

\begin{figure}[ht]
\begin{center}
\includegraphics[scale=0.45,angle=0]{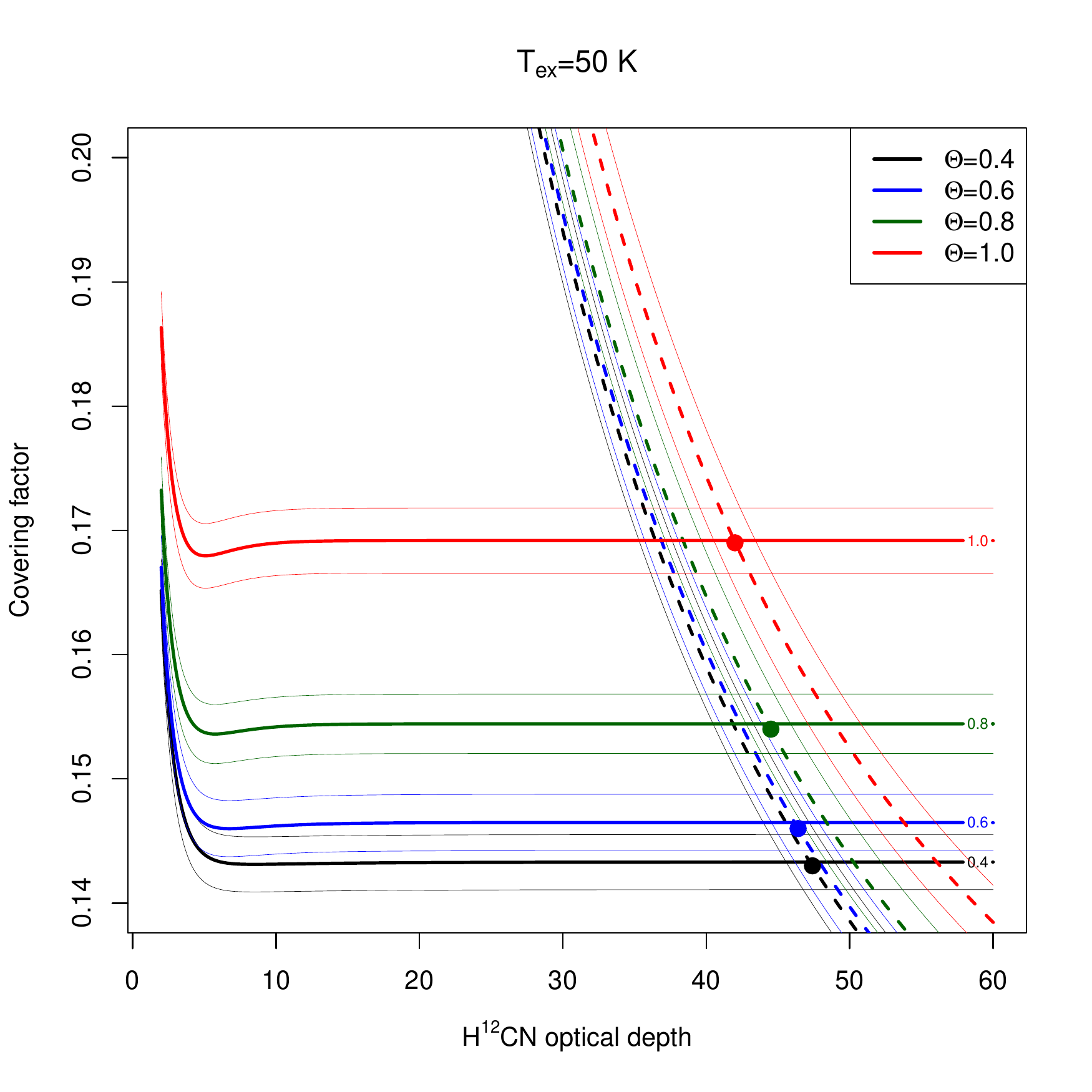}
\includegraphics[scale=0.45,angle=0]{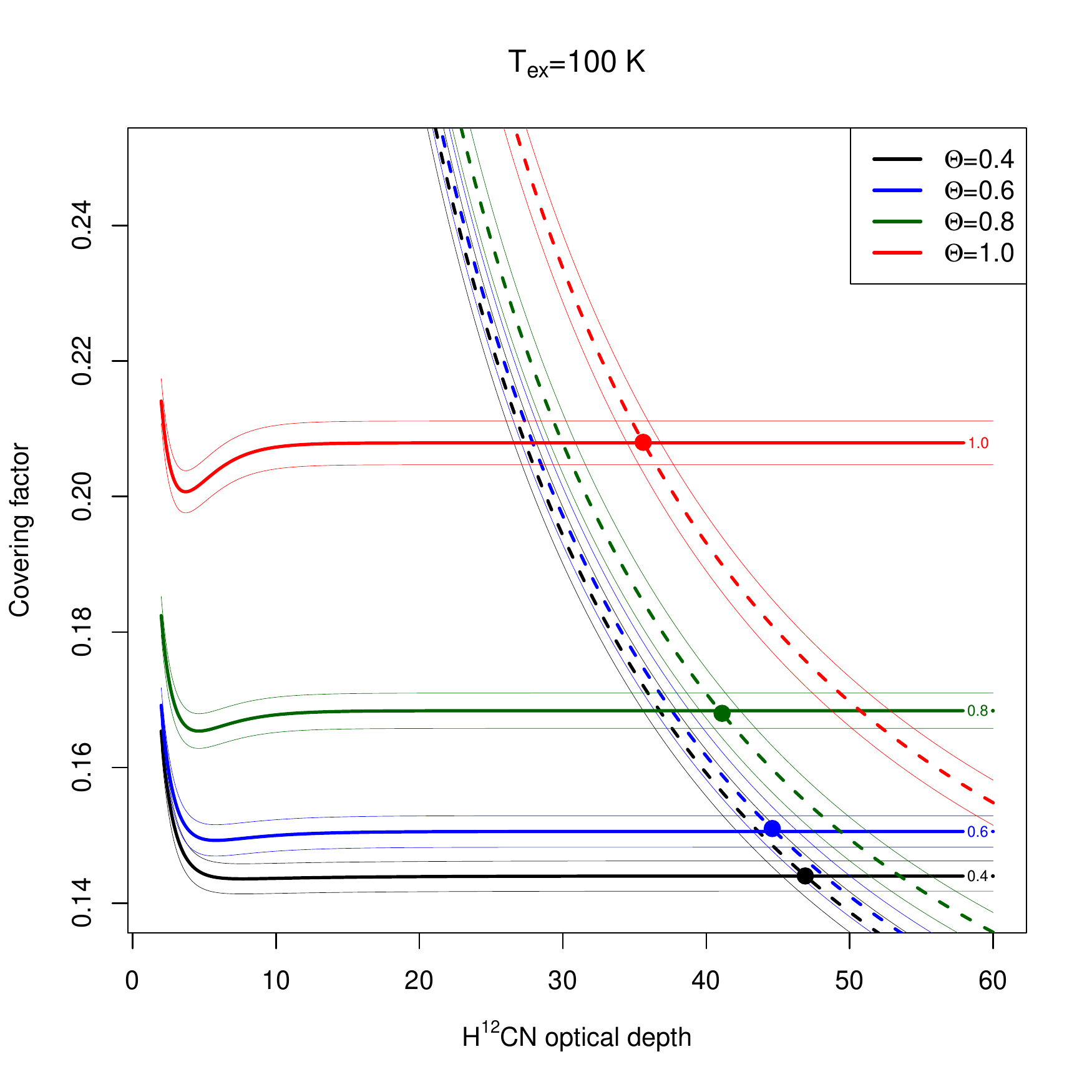}

\includegraphics[scale=0.45,angle=0]{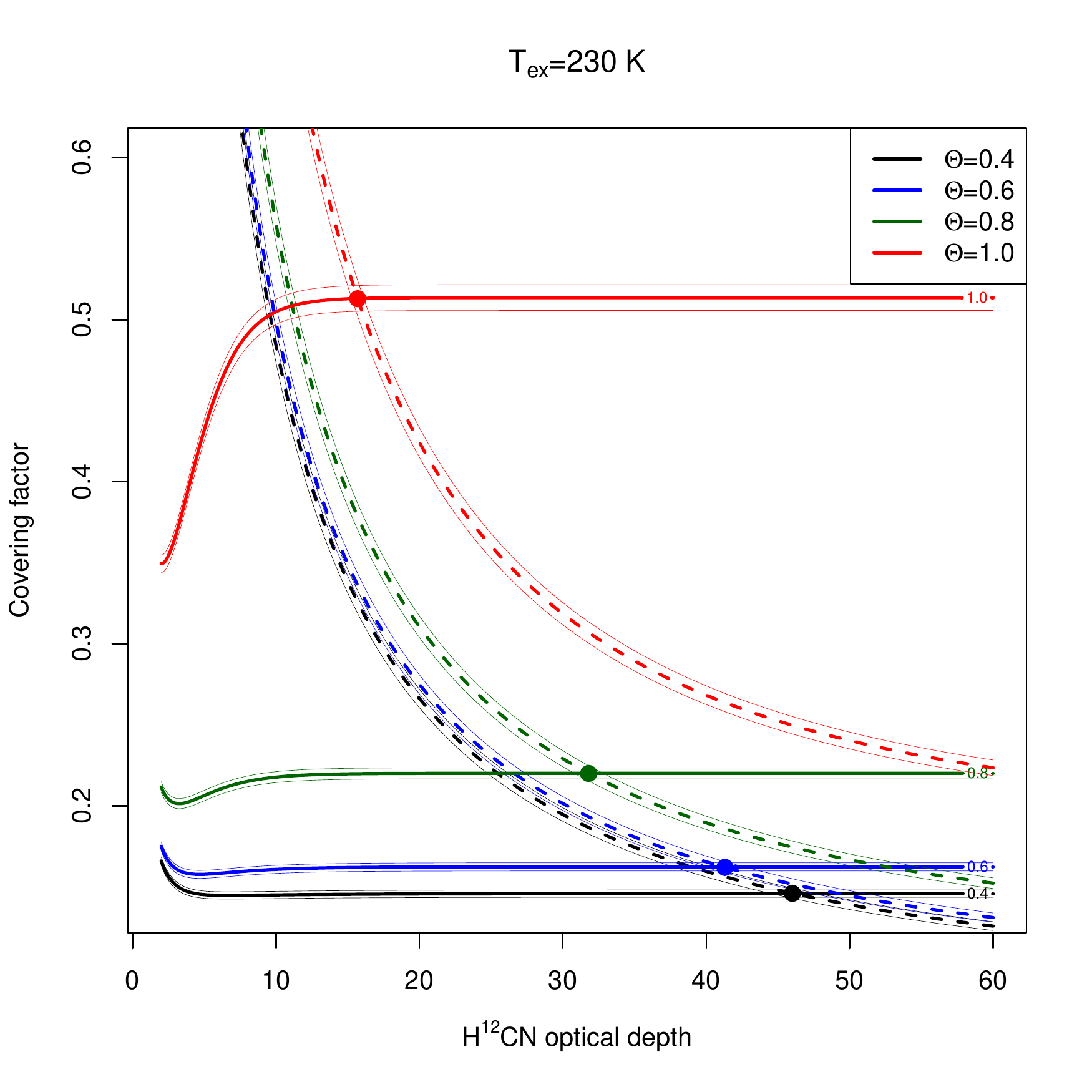}
\includegraphics[scale=0.45,angle=0]{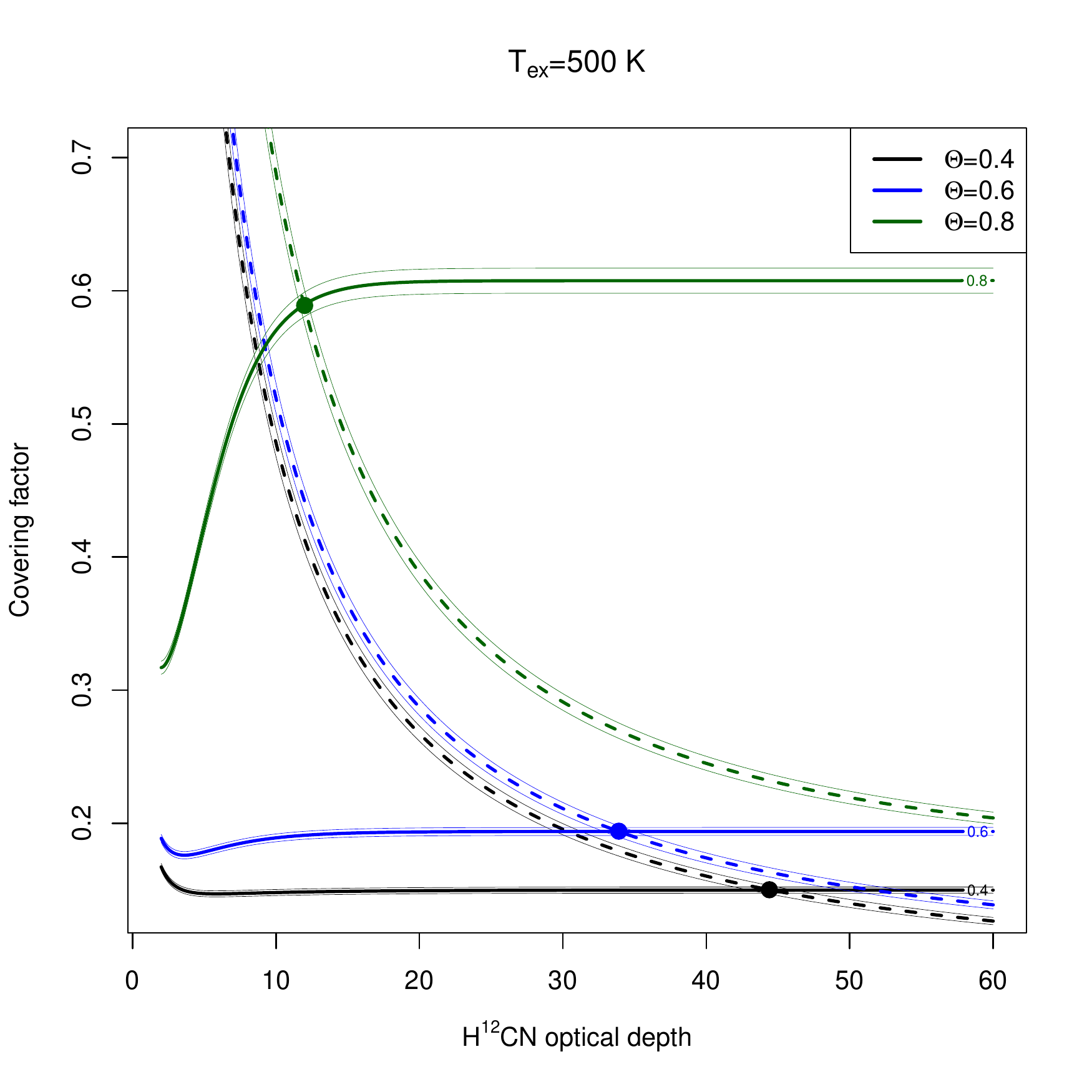}
\caption{Solutions of an optical depth and a covering factor for H$^{12}$CN and H$^{13}$CN flux densities modeled in equation \ref{eqn:radiationTransfer}, assuming $T_{\rm ex} = $ 50 K, 100 K, 230 K, and 500 K. Note that the range of the covering factor is optimized for each plot. Conditions for flux densities of H$^{12}$CN and H$^{13}$CN are drawn in solid and dashed lines, respectively. Thin lateral lines stand for standard errors of the observed flux densities. Colors stand for different torus thickness, $\Theta$. The cross points marked by filled circles indicate the solutions that jointly meet observed H$^{12}$CN and H$^{13}$CN flux densities. The parameters for the solutions are listed in table \ref{tab:fcovTau}}.
\label{fig:FcovTau}
\end{center}
\end{figure}

\begin{deluxetable}{llrrrrr}
\tablecaption{Solutions for the covering factor and the optical depth}
\tablehead{
\colhead{$T_{\rm ex}$} & \colhead{$\Theta$} & \colhead{$R$} & \colhead{$V_{\rm R}$} & \colhead{$f_{\rm cov}$}   & \colhead{$\tau_{\ell}$(H$^{12}$CN)}   & \colhead{$E_{\ell}$(H$^{12}$CN)}  \\
\colhead{(K)}     &  & (pc) & \colhead{(km s$^{-1}$)} &  &  & \colhead{(Jy)} 
 }
 \colnumbers
 \startdata
 50	  & 0.4 & $1.02$ & $1875$ & $0.143 \pm 0.001$  & $46.9 \pm 1.1$ & $ 3.0\times 10^{-4} $ \\
	  & 0.6 & $1.96$ & $ 833$ & $0.146 \pm 0.001$  & $46.2 \pm 0.8$ & $ 1.7\times 10^{-3} $ \\
	  & 0.8 & $2.89$ & $ 469$ & $0.154 \pm 0.001$  & $44.3 \pm 0.7$ & $ 5.2\times 10^{-3} $ \\
	  & 1.0 & $3.70$ & $ 300$ & $0.169 \pm 0.001$  & $42.4 \pm 0.5$ & $ 1.2\times 10^{-2} $ \\ \hline
100	  & 0.4 & $1.02$ & $1875$ & $0.144 \pm 0.001$  & $46.9 \pm 1.1$ & $ 6.1\times 10^{-4} $ \\
	  & 0.6 & $1.96$ & $ 833$ & $0.151 \pm 0.001$  & $44.8 \pm 0.8$ & $ 3.5\times 10^{-3} $ \\
	  & 0.8 & $2.89$ & $ 469$ & $0.168 \pm 0.001$  & $41.0 \pm 0.7$ & $ 1.1\times 10^{-2} $ \\
	  & 1.0 & $3.70$ & $ 300$ & $0.208 \pm 0.001$  & $35.7 \pm 0.7$ & $ 2.9\times 10^{-2} $ \\ \hline
230	  & 0.4 & $1.02$ & $1875$ & $0.146 \pm 0.001$  & $46.0 \pm 1.1$ & $ 1.4\times 10^{-3} $ \\
	  & 0.6 & $1.96$ & $ 833$ & $0.162 \pm 0.001$  & $41.4 \pm 1.2$ & $ 8.8\times 10^{-3} $ \\
	  & 0.8 & $2.89$ & $ 469$ & $0.220 \pm 0.001$  & $31.0 \pm 0.4$ & $ 3.4\times 10^{-2} $ \\
	  & 1.0 & $3.70$ & $ 300$ & $0.514 \pm 0.003$  & $15.7 \pm 0.2$ & $ 1.6\times 10^{-1} $ \\ \hline
500	  & 0.4 & $1.02$ & $1875$ & $0.150 \pm 0.001$  & $44.4 \pm 1.0$ & $ 3.2\times 10^{-3} $ \\
	  & 0.6 & $1.96$ & $ 833$ & $0.194 \pm 0.001$  & $33.9 \pm 0.7$ & $ 2.3\times 10^{-2} $ \\
	  & 0.8 & $2.89$ & $ 469$ & $0.590 \pm 0.003$  & $12.0 \pm 0.2$ & $ 2.0\times 10^{-1} $ \\ \hline
\enddata
\tablecomments{
(1) Excitation temperature; (2) Torus thickness; (3) Radius; (4) Rotation velocity; (5) Covering factor; (6) H$^{12}$CN optical depth of clamps; (7) Flux density of H$^{12}$CN emission. See also figure \ref{tab:fcovTau} for the solutions. No realistic solution ($f_{\rm cov} < 1.0$) meets the condition of $T_{\rm ex} = 100$ K and $\Theta = 1.0$.
}
\end{deluxetable} \label{tab:fcovTau}

\subsubsection{Covering factor and optical depth} \label{subsec:coveringFactor}
We found an abnormal isotopologue ratio of $N_{\rm tot}({\rm H}^{12}{\rm CN}) / N_{\rm tot}({\rm H}^{13}{\rm CN}) = 2.16 \pm 0.10$, which was too much different from the C$^{12}$/C$^{13}$ isotope ratio of $41.6 \pm 0.2$ \citep{2019A&A...629A...6T} or $21 \pm 6$ \citep{2019A&A...624A.125M} in the starburst galaxy NGC 253, $> 30$ in interstellar media \citep{1967ApJ...150..729A}, $43\pm 4$ in the solar neighborhood \citep{1987ApJ...317..926H}, or $67.5 \pm 1.0$ in diffuse molecular clouds \citep{2011ApJ...728...36R}.
If we assume that the typical C$^{12}$-to-C$^{13}$ abundance ratio of $\sim 50$ and that H$^{13}$CN absorption is optically thin, the expected peak optical depth of H$^{12}$CN would be $\sim 3.0$. Thus, the H$^{12}$CN absorption feature would be optically thick and the derived column density is underestimated.

Smaller number of the observed $\tau_{\rm max}$(H$^{12}$CN) than the expected optical depth can be explained by a small covering factor, $f_{\rm cov}$, of the absorber and/or dilution by line emission from the molecular torus.
Here we attempt modeling the molecular torus consisting of many small clumps (see figure \ref{fig:clumpyTorus}) and estimating the optical depth and the covering factor that meet the observed absorption features of H$^{12}$CN and H$^{13}$CN.
The flux density, $S_{\ell}$, at the bottom of an absorption line feature will be given as
\begin{eqnarray}
S_{\ell} = S_{c} (1 - f_{\rm cov}) + S_{c} f_{\rm cov} e^{-\tau_{\ell}} + E_{\ell}, \label{eqn:radiationTransfer}
\end{eqnarray}
where $S_{c}$ is the continuum flux density, $\tau_{\ell}$ is the optical depth of a clump, and $E_{\ell}$ is the summation of line emission from all clumps. Three terms in equation \ref{eqn:radiationTransfer} stand for unabsorbed continuum radiation, absorbed continuum radiation, and line emission. The flux density of line emission, $E_{\ell}$, is given as
\begin{eqnarray}
E_{\ell} = \frac{2 k T_{\rm ex} \Omega}{\lambda^2}  \left( 1 - \exp \left[ - \frac{\tau_{\ell}}{1 + \Theta^{-2}} \right] \right), \label{eqn:emissionFlux}
\end{eqnarray}
where $\lambda$ is the line rest wavelength.
While all clumps in the torus contribute to emission, only those in front of the continuum source can contribute to absorption.
Thus, line emission should have wider velocity width that dilutes the optical depth by a factor of $1 + \Theta^{-2}$.
If the torus is seen edge-on, the total solid angle of the clumps, $\Omega$, is given by $\Omega = 8 f_{\rm cov} R H / d^2_{\rm A}$ where $R$ and $H$ are radius and height of the torus, respectively.
Applying equation \ref{eqn:virialRadius}, we have
\begin{eqnarray}
\Omega = \frac{8 f_{\rm cov} r^2_{\rm vir} \Theta}{d^2_A (1 + \Theta^{-2})^2}. \label{eqn:solidAngle}  
\end{eqnarray}
Since the molecular torus is unresolved, the velocity width of the emission line corresponds to $\left< V^2_H \right>$ and thus the optical depth in equation \ref{eqn:emissionFlux} is diluted by the factor of $\left< V^2 \right> / \left< V^2_H \right> = 1 + \Theta^{-2}$. Note that the covering factor of line emission is doubled to include the backside of the core.

Then, we put constraints: $S_{\ell}$(H$^{12}$CN)$=0.3815 \pm 0.00033$ Jy,
$S_{\ell}$(H$^{13}$CN)$=0.4036 \pm 0.00028$ Jy, $S_{c}$(H$^{12}$CN)$=0.4440$ Jy, $S_{c}$(H$^{13}$CN)$=0.4423$ Jy, and $\tau_{\ell}($H$^{12}$CN)$ / \tau_{\ell}($H$^{13}$CN) $= 50$.
We also have $\Theta \geq 0.4$ if we assume that the molecular torus is larger than the plasma torus.
The solutions of $f_{\rm cov}$ and $\tau_{\ell}$ to meet the constraints with a given value of $\Theta=$ 0.4, 0.6, 0.8,and 1.0 are shown in figure \ref{fig:FcovTau} with the assumed excitation temperatures of 50 K, 100 K, 230 K, and 500 K. Obtained solutions of the H$^{12}$CN optical depths and the covering factor are listed in table \ref{tab:fcovTau} together with the estimated H$^{12}$CN emission flux density.
We found that $\Theta > 0.8$ and $\Theta > 1.06$ results in unrealistic solutions of $f_{\rm cov} > 1$ under $T_{\rm ex} = 500$ K and 230 K, respectively. Thus, the parameter range of $\Theta$ in $0.4 - 1.0$ for sufficiently covers a realistic torus model (except the case of $T_{\rm ex} = 500$ K) and that yields $R = 2.4 \pm 1.3$.

The results indicate very high H$^{12}$CN optical depth of $12.0 - 46.9$. The covering factor remains $f_{\rm cov} = 0.17^{+0.06}_{-0.03}$, except the case of high $T_{\rm ex}$ and large $\Theta$.
Let us compare with the KVN results \citep{2016ApJ...830L...3S} where H$^{12}$CN absorber is spatially resolved and intensity-weighted mean optical depths for two velocity components (I and II) are estimated as $\left< \tau \right> = 0.027$ and 0.028, respectively. Then, the covering factor of them are $f_{\rm cov} =  \left< \tau \right>_{\rm I} / \tau_{\rm max, \ I} + \left< \tau \right>_{\rm II} / \tau_{\rm max, \ II} = 0.17$.
Most of our $f_{\rm cov}$ estimation, except the case of high $T_{\rm ex}$ and large $\Theta$, are consistent with that obtained by the KVN.
If we omit the exception of high $T_{\rm ex}$ and large $\Theta$ case, H$^{12}$CN optical depth converges in a range of $\tau_{\ell}($H$^{12}$CN$) = 41^{+7}_{-10}$.
That requires optical depth of H$^{13}$CN $\sim 0.8$ and the assumption of optically thin H$^{13}$CN is no longer applicable.

The optical depths of abundant line species require a correction by the factor of $\tau_{\ell}($CO$_{3-2}) = 31 \pm 4$ and $\tau_{\ell}($HCO$^{+}_{4-3}) = 24 \pm 3$.
If we apply $f_{\rm cov} = 0.17$ and corrected optical depth to equation \ref{eqn:radiationTransfer} for abundant molecules of CO ($J=3-2$) and H$^{12}$CO$^{+}$, the flux density at the line bottom is estimated to be 0.379 Jy and 0.380 Jy, respectively, in any $\Theta$.
Those meet with the observed flux densities of $0.386 \pm 0.015$ Jy and $0.399 \pm 0.016$ Jy, respectively.

Corrected column densities of these line species are added in the last column of table \ref{tab:absorption}.
While the column density of H$^{13}$CN is multiplied by a factor of $\sim 2$, those of abundant molecules are magnified by a factor of $\sim 30 - 40$.

Although combination of H$^{12}$CN and H$^{13}$CN absorption lines allow as to estimate the covering factor and optical depths, a degeneracy between the optical depths and the torus thickness still remain.
This can be solved if we would be able to measure flux densities of line emission, which is sensitive to the torus thickness as shown in the last column of table \ref{tab:fcovTau}.
Future Long-baseline observations with ALMA would offer a capability to discriminate emission line of the molecular torus from absorption features and would allow us to determine the thickness.

\subsubsection{Vibrationally-excited absorption lines} \label{subsec:vib}
Three vibrationally-excited (v$_2=1$) absorption lines in HCN and HCO$^{+}$ were identified.
The velocity width of HCO$^{+}$ $J=4-3 \ {\rm v}_2=1, \ell=1f$ shows almost the same LSR velocity and FWHM with those in the HCN and HCO$^{+}$ $v=0$ line.
Thus, we consider that the vibrationally-excited absorption lines associates with the same pc-scale molecular torus.

The HCN v$_2=1$ emission line was found in luminous infrared galaxies such as NGC 4418 \citep{2010ApJ...725L.228S}, IRAS 20551-4250 \citep{2013AJ....146...91I}, and Arp 220 \citep{2016A&A...590A..25M}, and AGNs such as Mrk 231 \citep{2015A&A...574A..85A}.
Our results are the first case, as far as we know, that vibrationally excited HCN and HCO$^{+}$ lines are detected in absorption lines in a radio galaxy with a bright radio continuum.
Moreover, this is the first time vibrationally excited HCO$^{+}$ is reported outside the Galaxy, which may point out to fundamental physico-chemical differences with the obscured nuclei toward where vibrationally excited HCN has been reported so far \citep{2015A&A...584A..42A, 2017ApJ...849...29I}.

If the excitation is induced by the thermal collision, it requires $\sim 1000$ K and the estimated column density in table \ref{tab:absorption} would be multiplied by 4, which results in too high column density and inferred molecular gas mass estimated in the next subsection.
The realistic excitation is supposed to be caused by pumping through HCN absorption of 14 $\mu$m infrared emission from warm dust \citep{2007ApJ...659..296L}.
Although we don't have a direct detection of dust emission in the molecular torus, the presence of SO and SO$_2$ absorption suggests that these molecules are formed in the surface of dusts in the torus and are detached via irradiation.
Because SiO feature is absent, the irradiation energy is not high enough to disrupt silicates in the dust core.

\subsubsection{Molecular gas mass of the torus} \label{subsec:torusMass}
We attempt to estimate the molecular gas mass in the torus in this section.
The column densities of H$^{12}$CN, HCO$^{+}$, and CO have been estimated in \S \ref{subsec:coveringFactor} and listed in table \ref{tab:absorption} with the corrected optical depth and the covering factor.

Applying abundance ratios of $(0.2-4.1) \times 10^{-9}$ in HCN-to-H$_2$ \citep{2014MNRAS.437.3159S}, $(2 - 3) \times 10^{-9}$ in HCO$^{+}$-to-H$_2$ \citep{2010A&A...518A..45L}, and $\sim 10^{-4}$ in CO-to-H$_2$ \citep{2013ARA&A..51..207B}, we obtain the H$_2$ column density of $N_{\rm H_2} = (6 - 130) \times 10^{25}$ cm$^{-2}$, $(3.3 \pm 0.7) \times 10^{25}$ cm$^{-2}$, and $(1.5 \pm 0.2) \times 10^{24}$ cm$^{-2}$, respectively.
The differences in estimated H$_2$ column density is caused by uncertainties in abundance ratios (mainly in HCN) CO emission contamination from CND, or potentially greater covering factor of CO than HCN due to different critical densities.
\citet{2010ApJ...721.1570H, 2013PASJ...65..100I} showed that chemical gas-phase reaction network in the pc-scale vicinity of AGN yields a high production efficiently of HCN sensitive to temperature above 100 K, while that of HCO$^{+}$ is less sensitive.
To keep simplicity, we adopt the H$_2$ column density estimated by HCO$^{+}$ because of relatively small uncertainty in the abundance ratio. That value is greater than $N_{\rm H_2} \sim 10^{24} - 10^{25}$ cm$^{-2}$ estimated by VLBI observations of HCO$^{+}$ $J=1-0$ absorption \citep{2016ApJ...830L...3S, 2019ApJ...872L..21S}.
The difference would be ascribed to the spatial distribution of clumps with a higher density at a closer position to the nucleus, which has greater dominance in the continuum emission in our ALMA observation at 350 GHz than in the KVN observation at 89 GHz.


The mean H$_2$ number density is derived as $n^{\rm mean}_{\rm H_2} = N_{\rm H_2} / R = (4.4 \pm 0.9) \times 10^6$ cm$^{-3}$, where $R \sim 2.4$ pc for $\Theta = 0.7$.
Applying the volume filling factor as $f^{3/2}_{\rm cov} \sim 0.07$, the hydrogen density in each clump will be $n^{\rm clump}_{\rm H_2} = n^{\rm mean}_{\rm H_2} / f^{3/2}_{\rm cov} = (6.4 \pm 1.3) \times 10^7$ cm$^{-3}$, which approaches to the excitation condition of H$_2$O maser.
This implies that a layer of the absorbers in the torus is adjacent to the outside of H$_2$O maser emitting region.

The total mass of the torus is estimated as $M_{\rm torus} = 2\pi \Theta R^2 N_{\rm H_2} \mu_{\rm H_2} = (1.3 \pm 0.3) \times 10^7$ M$_{\odot}$, where $\mu_{\rm H_2}$ is the mass of a hydrogen molecule.
The estimated gas mass in the torus is $\sim 9$\% of the black-hole mass, and accounts for $\sim 0.6$\% of the enclosed mass inside the CND.
$R$ has an uncertainty by a factor of $\sim 2$ coupled with $\Theta$ as discussed in \S \ref{subsec:coveringFactor}, and that propagates to the mean H$_2$ density and th torus mass.
We have $n^{\rm mean}_{\rm H_2}$ of $(1.0 \pm 0.2) \times 10^7$ cm$^{-3}$ and $(2.9 \pm 0.6) \times 10^6$ cm$^{-3}$, and $M_{\rm torus}$ of $(1.4 \pm 0.3) \times 10^6$ M$_{\odot}$ and $(4.4 \pm 1.0) \times 10^7$ M$_{\odot}$ for ($\Theta = 0.4$ and $R=1.04$ pc) and ($\Theta = 1.0$ and $R=3.7$ pc) cases, respectively.

It is remarkable that the amount of gas in the torus is significantly greater than that in the CND.
This indicates that accretion from the CND onto the torus in NGC 1052 is currently not working.
Comparing with the Seyfert galaxy NGC 1068 where the molecular gas mass in the CND and the torus are estimated to be $1.4 \times 10^8$ M$_{\odot}$ and $1.2 - 1.8 \times 10^5$ M$_{\odot}$, respectively \citep{2019A&A...632A..61G}, we find a quite different situation in NGC 1052 that host a massive molecular torus inside a gas-poor CND.

\begin{figure}[ht]
\begin{center}
\includegraphics[scale=0.8,angle=0]{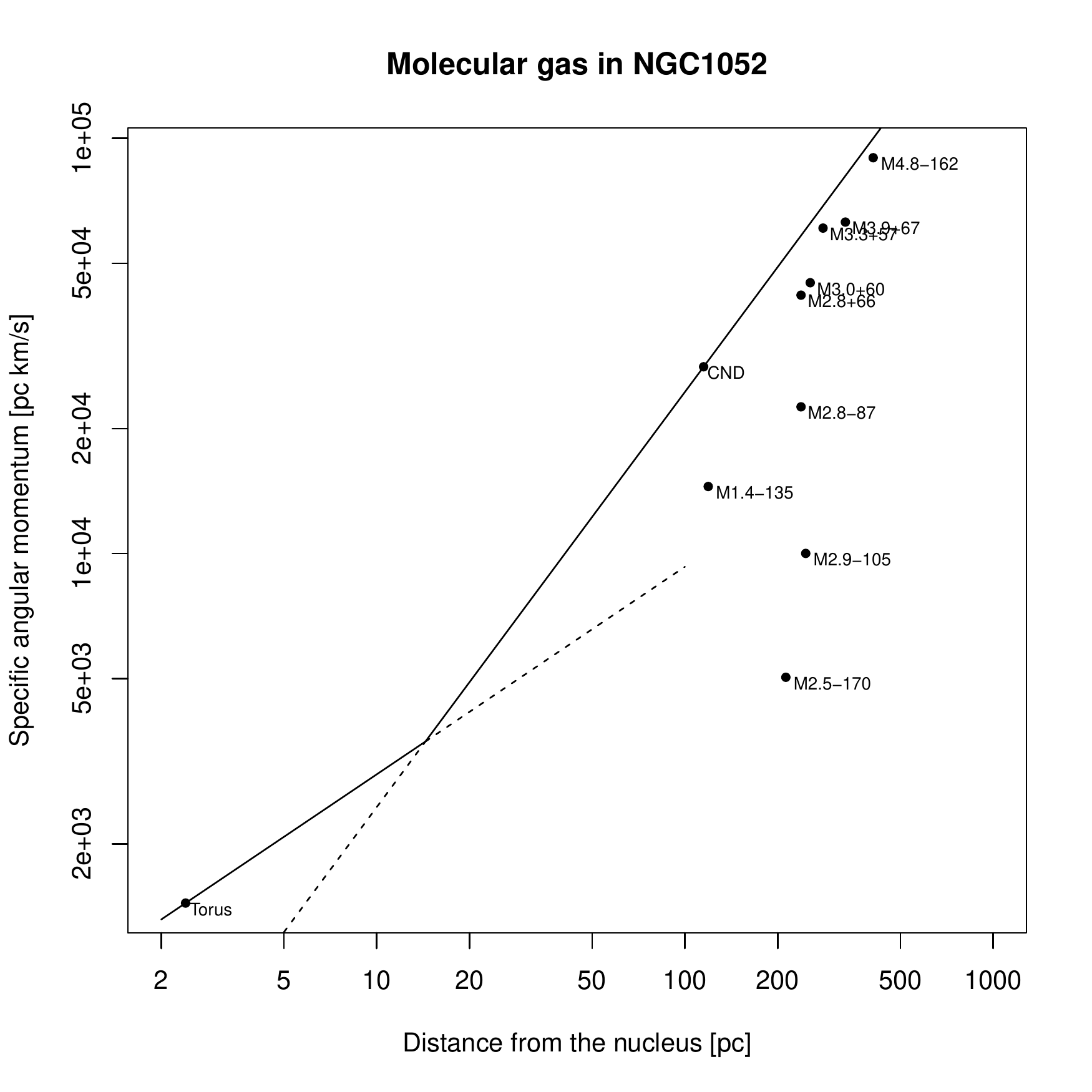}
\caption{Specific angular momentum (SAM) of gas clouds in NGC 1052. SAMs for  molecular clouds indicate lower limits because they are derived using the projected distances from the nucleus and the radial velocities. SAMs for the CND and the torus employ the radii of 115 pc and 2.4 pc, and the rotation velocities of 245 km s$^{-1}$ and 600 km s$^{-1}$, respectively. Lines passing the points of the CND and the torus stand for a flat rotation curve (SAM $\propto R$) and a Keplerian rotation (SAM $\propto R^{1/2}$), respectively. The cross point locates at $R=14.4$ pc.}
\label{fig:angularMomentum}
\end{center}
\end{figure}

\subsection{Angular momentum of the CND and the torus} \label{subsec:angularMomentum}
Let us trace the specific angular momentum (SAM) of gas in the torus, the CND, and the molecular clouds outside the CND.
Estimation of the rotation velocity in the torus is $V_{R} = 600$ km s$^{-1}$ for $\Theta = 0.7$ and that yields SAM $= V_R R = 1.4 \times 10^3$ pc km s$^{-1}$ for $R=2.4$ pc\footnote{The rotation yields the velocity gradient of $10^4$ km s$^{-1}$ arcsec$^{-1}$ and results in additional velocity width of $< 60$ km s$^{-1}$ toward the continuum source size less than $5.7 \times 5.1$ milliarcsec. This does not significantly affect discussion about the size of molecular torus.}.
Comparing with the CND (SAM $=2.8 \times 10^4$ pc km s$^{-1}$) and the molecular clouds (SAM $= (0.5 - 9.0) \times 10^4$ pc km s$^{-1}$) listed in table \ref{tab:MCflux}, we find that SAM of the torus is smaller by an order of magnitude as shown in figure \ref{fig:angularMomentum}. Note that SAM of molecular clouds in the plot are lower limits because they are derived by the projected radii from the nucleus and the radial velocities.

While SAMs of the molecular clouds and the CND follows a flat rotation curve as SAM $\propto R$, the SAM of the torus does not.
If we assume Keplerian rotation (SAM $\propto R^{1/2}$) at the radius of the torus and draw an outer extrapolation, it crosses the inward flat-rotation at $R=14.4$ pc, which indicates the border of the sphere of influence (SoI) and the galactic rotation.
Higher spatial resolution and sensitivity is required to clarify the gas dynamics at the border.
%
%


\section{Conclusions} \label{sec:conclusions}
We have carried out ALMA observations addressing circumnuclear region of the radio galaxy NGC 1052 to quest mass accretion onto the AGN.
Our findings are summarized below:
\begin{enumerate}
\item The nucleus is detected as an unresolved continuum core, with the size upper limit of $0.48 \times 0.43$ pc. The spectral index of $\alpha = -0.22 \pm 0.01$ between 222 GHz and 350 GHz is consistent with mm-VLBI observations that revealed a core-jet structure.

\item The circum-nuclear disk (CND) is discovered by CO $J=2-1$ and $J=3-2$ emission lines. The CND forms a rotating ring seen edge-on which the radius of 115 pc and the rotating velocity of $245 \pm 4.9$ km s$^{-1}$. The molecular gas mass of the CND is estimated as $M^{\rm CND}_{\rm H_2} = 5.3 \times 10^{5}$ M$_{\odot}$, which is two orders of magnitude smaller than CNDs in typical Seyfert galaxies. The CND gas mass is too little to trigger star formation that would drive mass accretion onto SMBH.
The enclosed mass inside the CND is measured to be $(2.13 \pm 0.09) \times 10^9$ M$_{\odot}$, which is significantly greater than the estimated SMBH mass of $1.5 \times 10^8$ M$_{\odot}$. Thus, the CND is supposed to reside outside of the sphere of influence (SoI).

\item The continuum spectrum casts a lot of absorption features such as CO, HCN, HCO$^+$, SO, SO$_2$, CS, CN, and H$_2$O. We have also detected isotopologues of H$^{13}$CN and HC$^{15}$N absorption features, and vibrationally excited ($v_2 = 1$) transitions of HCN and HCO$^+$. We conclude that these absorption features arise in a molecular torus surrounding the nucleus.

\item Presence of vibrationally excited absorption lines indicates that the molecular gas in the torus is warm and infrared pumping by 14$\mu$m continuum is working.

\item CO, HCN, and HCO$^+$ absorption lines are optically thick and then their optical depths are underestimated with a small covering factor of $f_{\rm cov} = 0.17^{+0.06}_{-0.03}$. Optically-thin H$^{13}$CN absorption feature allows us to estimate the H$_2$ column density of $N_{\rm H_2} = (3.3 \pm 0.7) \times 10^{25}$ cm$^{-2}$ across the torus toward the nucleus. This is slightly greater than the VLBI measurements of HCO$^{+}$ absorption \citep{2019ApJ...872L..21S}.

\item We have estimated gas densities averaged in the torus, $n^{\rm mean}_{\rm H_2} = (4.4 \pm 0.9) \times 10^6$ cm$^{-3}$, and in the clumps, $n^{\rm clump}_{\rm H_2} = (6.4 \pm 1.3) \times 10^7$ cm$^{-3}$.
The gas density in the clumps approaches to the excitation condition of the H$_2$O masers.
This indicates that the layer of the absorption lines resides adjacent to the H$_2$O maser emitting region.

\item The total gas mass of the molecular torus is estimated as $M_{\rm torus} = (1.3 \pm 0.3) \times 10^7$ M$_{\odot}$, which corresponds to $\sim 9$\% of the SMBH mass and accounts for $\sim 0.6$\% of the enclosed mass inside the CND. NGC 1052 hosts a massive molecular torus inside a gas-poor CND.

\item The specific angular momentum (SAM) of the molecular torus is smaller than those of the CND and the molecular clouds by an order of magnitude, and exceeds the extrapolation of a flag-rotation curve which the CND and the molecular clouds follow.
If we draw a Keplerian rotation curve that the torus follows, we find the border of the SoI at the radius of 14.4 pc.

\end{enumerate}

Our study demonstrates that millimeter and submillimeter obseravations of molecular emission/absorption lines toward neaby radio galaxies contribute to understand mass accretion process in pc-scale vicinity of a central engine.
Higher-resolution observations are desired to distinguish the molecular torus emission from the nuclear continuum to determine the geometry and dynamics of the torus.

\bigskip
We thank Dr. Swara Ravindranath for consent to extract the HST image used in figure \ref{fig:HST+CO}.
S. K. is supported by JSPS KAKENHI grant number 18K0371 and the ALMA Japan Research Grant of NAOJ Chile Observatory, NAOJ-ALMA-0116.
S. S.-S. is supported by Yamaguchi University Faculty of Science Stepup Research Grant 2019. 
K. K. is supported by JSPS KAKENHI Grant Number JP17H06130 and NAOJ ALMA Scientific Research Grant Number 2017-06B.
This paper makes use of the following ALMA data: ADS/JAO.ALMA\#2013.1.01225.S. ALMA is a partnership of ESO (representing its member states), NSF (USA) and NINS (Japan), together with NRC (Canada), MOST and ASIAA (Taiwan), and KASI (Republic of Korea), in cooperation with the Republic of Chile. The Joint ALMA Observatory is operated by ESO, AUI/NRAO and NAOJ. 

\facility{ALMA}

\software{CASA 5.6.0 (McMullin et al. 2007), R 3.6.1 (The R Foundation)}



\appendix

\section{CO Channel Maps} \label{appendix:channmap}

\begin{figure}[ht]
\begin{center}
\includegraphics[scale=0.85,angle=0]{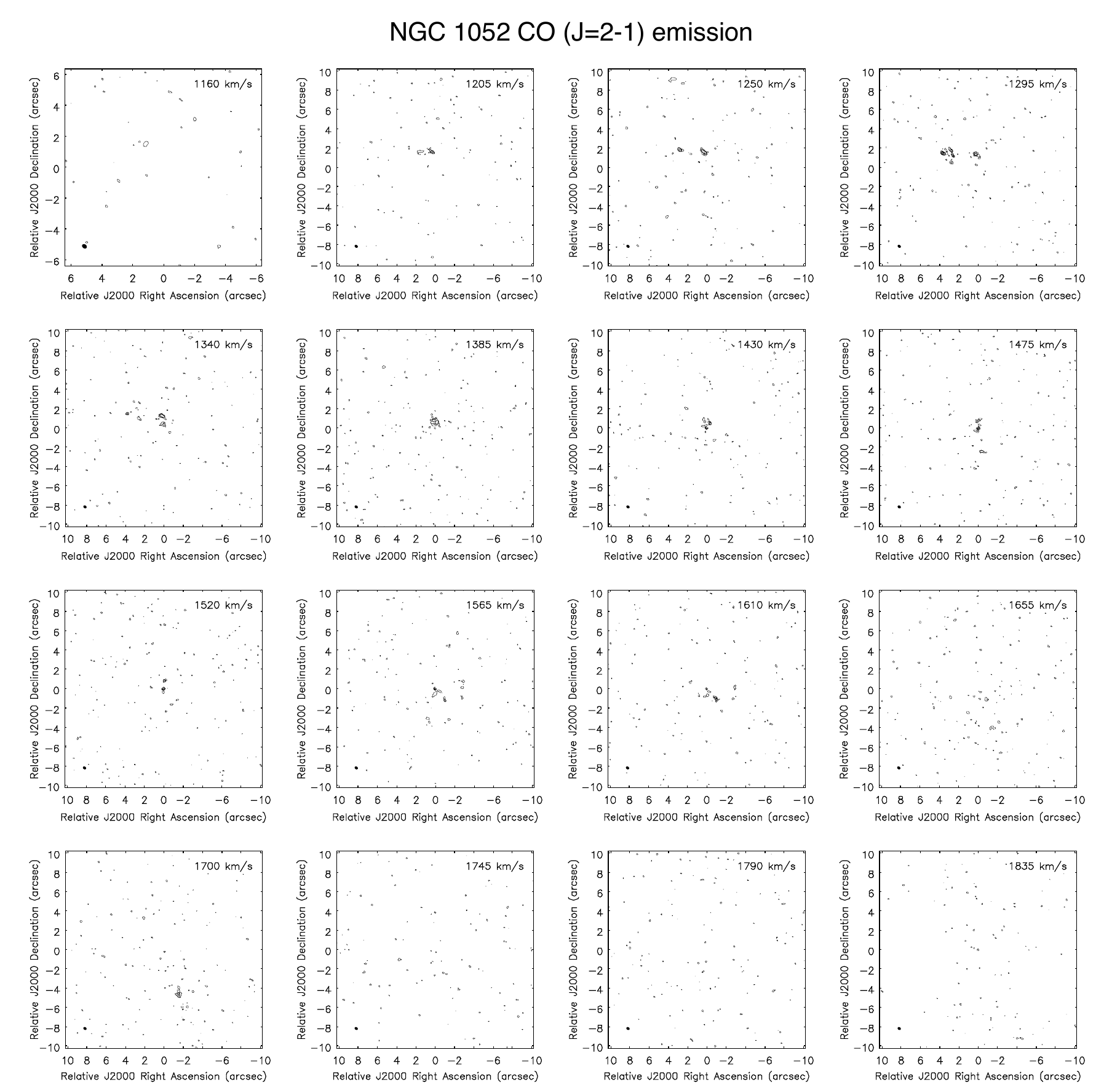}
\caption{Channel map of CO ($J=2-1$) emission with a velocity width of 45 km s$^{-1}$. Visibilities are tapered at 1 M$\lambda$ to match the resolution with the CO $J=3-2$ map in figure \ref{fig:CO32ChanMap}. The contours are at $2^n \times 0.76$ mJy beam$^{-1}$, 
where $n$ is integer. \label{fig:CO21ChanMap}}
\end{center}
\end{figure}

\begin{figure}[ht]
\begin{center}
\includegraphics[scale=0.85,angle=0]{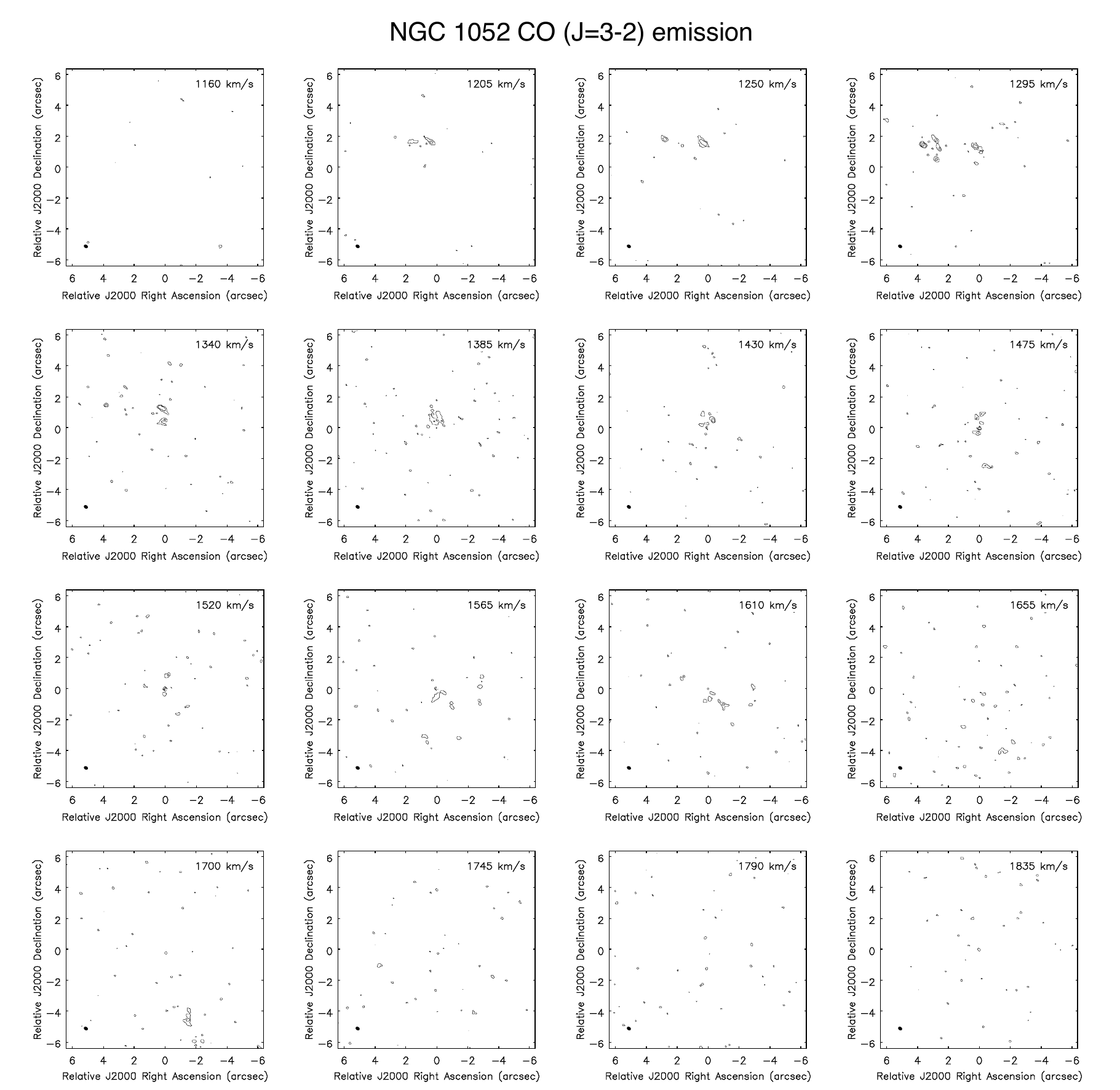}
\caption{Channel map of CO ($J=3-2$) emission. The velocity width and $(u, v)$ tapering are set the same with CO $J=2-1$ map in figure \ref{fig:CO21ChanMap}. The contours are at $2^n \times 1$ mJy beam$^{-1}$, where $n$ is integer.\label{fig:CO32ChanMap}}
\end{center}
\end{figure}

Channel maps of CO $J=2-1$ and $J=3-2$ emissions with a velocity width of 45 km s$^{-1}$ are shown in figures \ref{fig:CO21ChanMap} and \ref{fig:CO32ChanMap}.

 \end{CJK*}
\end{document}